\documentclass[a4paper,11pt]{article}
\usepackage[left=2cm,right=2cm,top=2cm,bottom=2cm]{geometry}
\usepackage{psfrag,amsmath,amsfonts,amssymb,latexsym,amsthm,verbatim,color,lscape}
\usepackage[figuresright]{rotating}
 \usepackage{pdfpages}
\usepackage{hyperref}
\hypersetup{colorlinks,
citecolor=black,
filecolor=black,
linkcolor=black,
urlcolor=black,
pdftex}
\usepackage{color}

\usepackage{graphicx}

\usepackage[super,sort&compress]{natbib}
\usepackage{amsmath}
\usepackage{algorithm}
\usepackage[noend]{algpseudocode}

\makeatletter
\def\BState{\State\hskip-\ALG@thistlm}
\makeatother

\usepackage{booktabs}

\usepackage{colortbl}
\definecolor{gray}{rgb}{0.8,0.8,0.8}

\usepackage{times}
\usepackage{blindtext}
\usepackage[font={footnotesize,it}]{caption}

\usepackage{sectsty}  
\subsectionfont{\normalsize\bf}
\sectionfont{\large\bf}

\def \red#1{\textcolor{red}{#1}}

\def \red#1{\textcolor{red}{#1}}

\begin{document}

\def\p{\partial}
\def\oo{\infty}
\def\rt#1{\sqrt{#1}\,}

\def\Cbar{{\overline C}}
\def\C{\mathbf{C}}
\def\E{{\rm E}\,}
\def\I{\mathbf{I}}
\def\pp{\mathbf{p}}
\def\R{\mathbf{R}}
\def\y{\mathbf{y}}
\def\Y{\mathbf{Y}}
\def\z{\mathbf{z}}
\def\x{\mathbf{x}}
\def\o{\omega}
\def\s{\sigma}

\def\V{\mathbf{V}}
\def\I{\mathbf{I}}
\def\bfv{\mathbf{v}}
\def\X{\mathbf{X}}
\def\D{\mathbf{D}}

\def\a{{\alpha}}
\def\g{\gamma}
\def\b{\beta}

\def\de{\delta}
\def\debf{\boldsymbol{\delta}}
\def\e{\epsilon}
\def\th{\theta}
\def\r{\rho}
\def\thbf{\boldsymbol{\theta}}
\def\taubf{\boldsymbol{\tau}}
\def\pibf{\boldsymbol{\pi}}
\def\Xibf{\boldsymbol{\Xi}}
\def\Sbf{\boldsymbol{\Sigma}}

\def \red#1{\textcolor{red}{#1}}
\def \blue#1{\textcolor{blue}{#1}}
\def \magenta#1{\textcolor{magenta}{#1}}
\def \green#1{\textcolor{green}{#1}}
\def\bbf{\boldsymbol{\beta}}
\def\bmu{\boldsymbol{\mu}}

\title{A D-vine copula mixed model for joint meta-analysis and comparison of  diagnostic tests}
\author{Aristidis K Nikoloulopoulos}

\date{}
\author{
Aristidis K. Nikoloulopoulos\footnote{{\small\texttt{A.Nikoloulopoulos@uea.ac.uk}}, School of Computing Sciences, University of East Anglia, Norwich NR4 7TJ, UK} }
\maketitle

\begin{abstract}
\baselineskip=24pt
\noindent For a particular disease there may be two  diagnostic tests developed, where each of the tests is subject to several studies. A quadrivariate  generalized linear mixed model (GLMM) has been recently  proposed to joint meta-analyse and compare two  diagnostic tests.
We propose a D-vine copula mixed model for joint meta-analysis and comparison of  two diagnostic tests.  
Our general model includes the quadrivariate GLMM as a special case and can also operate on the original scale of sensitivities and  specificities.
The method allows the direct calculation of sensitivity and specificity for each test, as well as, the parameters of the summary receiver operator characteristic (SROC) curve, along with  a  comparison between the SROCs of each test. Our  methodology is demonstrated with an extensive simulation study and illustrated by meta-analysing two examples where 2 tests for the diagnosis of a particular disease are compared. Our study suggests that there can be an improvement on GLMM in fit  to data since our model can also provide tail dependencies and asymmetries. \\

\end{abstract}

\section{Introduction}
\baselineskip=25pt
So far, work on multivariate  methods and models for meta-analysis of  diagnostic studies has mainly focused on a single test. \cite{JacksonRileyWhite2011,MavridisSalanti13,Ma-etal-2013} However, for a particular disease there may be two (or more) diagnostic tests developed, where each of the tests is subject to several studies; e.g., Takwoingi et al. \cite{Takwoingi-etal-2013} found a considerable large number of  systematic reviews which compared the accuracy of two (or more) tests.
Hence, one may want to combine all such studies to see how the competing tests are performing with respect to each other, and choose the best for clinical practice. \citep{Siadaty&Shu2004,Tatsioni-etal-2005,Leeflang-etal-2008}

To compare the accuracy of two (or more) tests  to a common gold standard, reviewers typically take one of two approaches: perform a separate bivariate meta-analysis, e.g., fit a generalized linear mixed model  \cite{Chu&Cole2006} (GLMM) to synthesize information for each test and then compare the meta-analytic summaries, or perform a bivariate GLMM meta-regression using the type of test as a categorical predictor. Both methods assume that all individuals studies are independently sampled (i.e., that the tests study two distinct sets of individuals, each containing individuals with disease and individuals without disease). But when the same individuals receive both tests, their results are correlated and one has to take into account any possible correlations between the 2 tests.  Because the two common approaches do not take these correlations into account, neither is a valid method for comparing tests performed on the same individuals. \citep{trikalinos-etal-2014-rsm}

A valid statistical model that accounts for this dependence, namely a quadrivariate GLMM, has been recently proposed for the comparison of two diagnostic tests.\cite{hoyer&kuss-2016-smmr} In this paper, we propose a vine copula mixed model as an extension of the quadrivariate GLMM by rather using a vine copula representation of the random effects distribution with normal and beta margins. Our general model (a) includes the quadrivariate GLMM as a special case, (b) can also operate on the original scale of sensitivities and  specificities, and (c) can also provide tail dependencies and asymmetries since the random effects distribution is expressed via vine copulas that allow for flexible dependence modelling, different from assuming simple linear correlation structures, normality and tail independence as in GLMM.
In fact, we extend the vine copula mixed model proposed by Nikoloulopoulos \cite{Nikoloulopoulos2015c} to the quadrivariate case. To do so, we employ a drawable vine (D-vine) which can nicely capture the dependence within and between the diagnostic tests.  

A  vine copula approach for meta-analysis for the comparison of two diagnostic tests was recently proposed by Hoyer and Kuss \cite{hoyer&kuss-2017-sim} who explored the use of a quadrivariate vine copula model for observed discrete variables (number of true positives and  true negatives for both tests) which have beta-binomial margins. This approach is actually an approximated likelihood  method for estimating a D-vine copula mixed model with beta margins for the latent vector of sensitivities and specificities.

The remainder of the paper proceeds as follows. 
Section \ref{copula-mixed-model-sec} has a brief overview of relevant copula theory and then introduces the D-vine copula mixed model for the comparison of two diagnostic tests and discusses its relationship with existing models.  Section \ref{miss-section} contains small-sample efficiency calculations to investigate the effect of misspecifying the random effects distribution on parameter estimators and standard errors  and compare the proposed methodology with existing methods.  Section \ref{app-sec} illustrates  our methodology with two examples where 2 tests for the diagnosis of a particular disease are compared. We conclude with some discussion in Section \ref{discussion}.

\section{\label{copula-mixed-model-sec}The vine copula mixed model for diagnostic test accuracy studies  }
In this section, we  introduce the D-vine copula mixed model for  the comparison of two diagnostic tests and discuss its relationship with existing  models. Before that, the first subsection has some background on copula models.
We complete this section with details on maximum likelihood estimation.

\subsection{\label{overview}Overview and relevant background for vine copulas}
A copula is a multivariate cdf with uniform $U(0,1)$ margins.\cite{joe97,joe2014,nelsen06}
If $F$ is a $d$-variate cdf with univariate margins $F_1,\ldots,F_d$,
then Sklar's \cite{sklar1959}  theorem implies that there is a copula $C$ such that
  $$F(x_1,\ldots,x_d)= C\Bigl(F_1(x_1),\ldots,F_d(x_d)\Bigr).$$
The copula is unique if $F_1,\ldots,F_d$ are continuous.
If $F$ is continuous and $(Y_1,\ldots,Y_d)\sim F$, then the unique copula
is the distribution of $(U_1,\ldots,U_d)=\left(F_1(Y_1),\ldots,F_d(Y_d)\right)$ leading to
  $$C(u_1,\ldots,u_d)=F\Bigl(F_1^{-1}(u_1),\ldots,F_d^{-1}(u_d)\Bigr),
  \quad 0\le u_j\le 1, j=1,\ldots,d,$$
where $F_j^{-1}$ are inverse cdfs \cite{nikoloulopoulos&joe12}. In particular,
if $\Phi_d(\cdot;\R)$
is the multivariate normal (MVN) cdf with correlation matrix $$\R=(\rho_{jk}: 1\le j<k\le d)$$ and
N(0,1) margins, and $\Phi$ is the univariate standard normal cdf,
then the MVN copula is
\begin{equation}\label{MVNcdf}
C(u_1,\ldots,u_d)=\Phi_d\Bigl(\Phi^{-1}(u_1),\ldots,\Phi^{-1}(u_d);\R\Bigr).
\end{equation}

A copula $C$ has reflection symmetry if  $(U_1,\ldots,U_d)\sim C$
implies that $(1-U_1,\ldots,1-U_d)$ has the same distribution $C$. This is apparently the case for the MVN copula. 
When it is necessary to have copula models with reflection asymmetry
and flexible lower/upper tail dependence, then vine copulas
are the best choice. \cite{joeetal10} The
$d$-dimensional vine copulas are built via successive mixing from
$d(d-1)/2$ bivariate linking copulas on trees and their cdfs
involve lower-dimensional integrals. Since the densities of multivariate
vine copulas can be factorized in terms of bivariate linking copulas
and lower-dimensional margins, they are computationally tractable.

\begin{figure}[!h]
\vspace{1cm}
\begin{center}
\setlength{\unitlength}{1.cm}
\begin{picture}(5,1)
\put(-0.3,0.7){\framebox(0.5,0.5){$1$}}
\put(0.2,1){\line(1,0){1}}
\put(0.7,1.2){\makebox(0,0){$12$}}
\put(1.2,0.7){\framebox(0.5,0.5){$2$}}
\put(1.7,1){\line(1,0){1}}
\put(2.2,1.2){\makebox(0,0){$23$}}
\put(2.7,0.7){\framebox(0.5,0.5){$3$}}
\put(3.2,1){\line(1,0){1}}
\put(3.7,1.2){\makebox(0,0){$34$}}
\put(4.2,0.7){\framebox(0.5,0.5){$4$}}

\put(7,0.9){$T_1$}

\put(0.5,-0.5){\framebox(0.5,0.5){$12$}}
\put(1.,-0.25){\line(1,0){1}}
\put(1.5,-0.){\makebox(0,0){\small$13|2$}}
\put(2.,-0.5){\framebox(0.5,0.5){$23$}}
\put(2.5,-0.25){\line(1,0){1}}
\put(3.,-0.){\makebox(0,0){\small$24|3$}}
\put(3.5,-0.5){\framebox(0.5,0.5){$34$}}

\put(7,-0.3){$T_2$}

\put(0.5,-2){\framebox(0.7,0.7){$13|2$}}
\put(1.2,-1.65){\line(1,0){1.5}}
\put(2.,-1.4){\makebox(0,0){\small$14|23$}}
\put(2.7,-2){\framebox(0.7,0.7){$24|3$}}

\put(7,-1.8){$T_3$}

\end{picture}
\end{center}
\vspace{2cm}
\caption{\label{4dvine}4-dimensional D-vine with 3 trees/levels
($T_j$, $j=1,\ldots,4$).}
\end{figure}
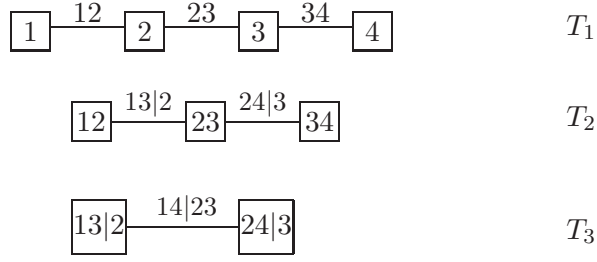

For the $d$-dimensional D-vine, the pairs at level 1 are $j,j+1$, for
$j=1,\ldots,d-1$, and for level $\ell$ ($2\le\ell<d$), the (conditional)
pairs are $j,j+\ell|j+1,\ldots,j+\ell-1$ for $j=1,\ldots,d-\ell$.
That is, for the D-vine, conditional copulas are specified
for variables $j$ and $j+\ell$ given the variables indexed in between \cite{nikoloulopoulos&joe&li11}. 
In Figure \ref{4dvine} 
a D-vine 
with 4 variables and 3 trees/levels is
depicted.

The quadrivariate D-vine density is decomposed in a simple manner by multiplying the nodes of the
nested set of trees
\begin{eqnarray}\label{vine-density}
  f_{1234}(x_1,x_2,x_3,x_4)&=&f_1(x_1)f_2(x_2)f_3(x_3)
  f_4(x_4)c_{12}\bigl(F_1(x_1),F_2(x_2)\bigr)\,c_{23}\bigl(F_2(x_2),F_3(x_3)\bigr)
\times \nonumber\\&&\,c_{34}\bigl(F_3(x_3),F_4(x_4)\bigr)  
 c_{13|2}\bigl(F_{1|2}(x_1|x_2),F_{3|2}(x_3|x_2)\bigr)c_{24|3}\bigl(F_{2|3}(x_2|x_3), \nonumber\\&&F_{4|3}(x_4|x_3)\bigr)c_{14|23}\bigl(F_{1|23}(x_1|x_2,x_3),F_{4|23}(x_4|x_2,x_3)\bigr)  \nonumber\\
&=&f_1(x_1)f_2(x_2)f_3(x_3)f_4(x_4)c_{1234}\bigl(F_1(x_1),F_2(x_2),F_3(x_3),F_4(x_4)\bigr),
\end{eqnarray}
where $F_{j|k}(x_j|x_k)=\partial C_{jk}\bigl(F_j(x_j),F_k(x_k)\bigr)/\partial F_k(x_k)$ \cite{joe96}.

\subsection{\label{normal-parametrization}The D-vine copula mixed  model for the latent pair of transformed sensitivity and specificity}
We first introduce the notation used in this paper. The focus is on two-level (within-study and between-studies) cluster data. 
The data  are $(y_{ij}, n_{ij}),\, i = 1, . . . ,N,\, j=1,2,3,4$, where $j$ is an index for the within study measurements and $i$ is an
index for the individual studies.
The data, for study $i$, can be summarized in two  $2\times 2$ tables with the number of true positives
($y_{i1}$), true negatives ($y_{i2}$), false negatives ($n_{i1}-y_{i1}$), and false positives ($n_{i2}-y_{i2}$) for the first diagnostic test, and the number of true positives
($y_{i3}$), true negatives ($y_{i4}$), false negatives ($n_{i3}-y_{i3}$), and false positives ($n_{i4}-y_{i4}$) for the second diagnostic test. We assume that the gold standard is the same for both tests, i.e., $n_{i1}=n_{i3}$ and $n_{i2}=n_{i4}$.

Here we generalize the quadrivariate  GLMM \cite{hoyer&kuss-2016-smmr}  by  proposing a model that links the four random effects using a D-vine copula  rather than the quadrivariate normal (QVN) distribution.
The within-study model assumes that the number of true positives $Y_{i1}$ and true negatives $Y_{i2}$ for the first test and the number of true positives $Y_{i3}$ and true negatives $Y_{i4}$ for the second test are conditionally independent and binomially distributed given $(\X_1,\X_2)=(\x_1,\x_2)$, where $\X_1=(X_1,X_2)$ denotes the  bivariate latent  pair of  (transformed)  sensitivity and specificity for the first test and $\X_2=(X_3,X_4)$ denotes the  bivariate latent  pair of (transformed) sensitivity and specificity for the second test.  
That is
\begin{eqnarray}\label{withinBinom}
Y_{i1}|X_{1}=x_1&\sim& \mbox{Binomial}\bigl(n_{i1},l^{-1}(x_1)\bigr);\nonumber\\
Y_{i2}|X_{2}=x_2&\sim& \mbox{Binomial}\bigl(n_{i2},l^{-1}(x_2)\bigr);\\
Y_{i3}|X_{3}=x_3&\sim& \mbox{Binomial}\bigl(n_{i1},l^{-1}(x_3)\bigr);\nonumber\\
Y_{i4}|X_{4}=x_4&\sim& \mbox{Binomial}\bigl(n_{i2},l^{-1}(x_4)\bigr),\nonumber
\end{eqnarray}
where $l(\cdot)$ is a link function.

The stochastic representation of the between studies model takes the form
\begin{equation}\label{copula-between-norm}
\Bigl(F\bigl(X_1;l(\pi_1),\de_1\bigr),F\bigl(X_2;l(\pi_2),\de_2\bigr),F\bigl(X_3;l(\pi_3),\de_3\bigr),F\bigl(X_4;l(\pi_4),\de_4\bigr)\Bigr)\sim C(\cdot;\thbf),
\end{equation}
where $C(\cdot;\thbf)$ is a D-vine  copula with dependence parameter vector $\thbf=(\th_{12},\th_{23},\th_{34},\th_{13|2},\th_{24|3},\th_{14|23})$ and $F\bigl(\cdot;l(\pi),\de\bigr)$ is the cdf of the univariate distribution of the random effect.

The models in (\ref{withinBinom}) and (\ref{copula-between-norm}) together specify a D-vine copula mixed  model with joint likelihood

\begin{multline}\label{mixed-cop-likelihood}
L(\pi_1,\pi_2,\pi_3,\pi_4,\de_1,\de_2,\de_3,\de_4,\thbf)=\\
\prod_{i=1}^N\int_{0}^{1}\int_{0}^{1}\int_{0}^{1}\int_{0}^{1}
\prod_{j=1}^4g\Bigl(y_{ij};n_{ij},l^{-1}\bigl(F^{-1}(u_j;l(\pi_j),\de_j)\bigr)\Bigr)c_{1234}(u_1,u_2,u_3,u_4;\thbf)du_1du_2du_3du_4, 
\end{multline}
 where 
$g\bigl(y;n,\pi\bigr)=\binom{n}{y}\pi^y(1-\pi)^{n-y},\quad y=0,1,\ldots,n,\quad 0<\pi<1,$
 is the binomial probability mass function (pmf).

 The copula parameter vector $\thbf$ has parameters of the random effects model and they are separated from the univariate parameters $(\pi_j,\de_j),\,j=1,\ldots,4$.  The parameters $\pi_1$ and $\pi_2$ are the meta-analytic parameters for the sensitivity and specificity,  $\de_1$ and $\de_2$  express the between-study variabilities for the first test, and
$\pi_3$ and $\pi_4$ are the meta-analytic parameters for the sensitivity and specificity,   $\de_3$ and $\de_4$  express the between-study variabilities for the second test.
 The choices of the  $F\bigl(\cdot;l(\pi),\de\bigr)$ and  $l$ are given in Table \ref{choices}.  If the Beta$(\pi,\gamma)$ distribution is used for the marginal modelling of
the latent proportions, then  
 one  does not have to transform the latent sensitivities and specificities and can work on the original scale. Our general statistical model allows for selection of copulas and margins independently, i.e., there are no constraints in the choices of parametric copulas and margins.

\setlength{\tabcolsep}{37pt}
\begin{table}[!h]
\begin{center}
\caption{\label{choices}The choices of the  $F\bigl(\cdot;l(\pi),\de\bigr)$ and  $l$ in the copula mixed model.}
\begin{tabular}{cccc}
\hline $F\bigl(\cdot;l(\pi),\de\bigr)$ & $l$ & $\pi$ & $\de$\\\hline
$N(\mu,\s)$ & logit, probit, cloglog & $l^{-1}(\mu)$&$\s$\\
Beta$(\pi,\gamma)$ & identity & $\pi$ & $\gamma$\\
\hline
\end{tabular}
\end{center}
\end{table}

\subsection{Relationship with existing models}

\subsubsection{Relationship with the quadrivariate GLMM}
In this subsection, we show what happens when all the bivariate copulas are bivariate normal (BVN)  and the univariate distribution of the random effects is the  $N(\mu,\s)$ distribution.  
One can easily deduce that the within-study model in (\ref{withinBinom}) is the same as in the quadrivariate GLMM. 

Furthermore, when all the bivariate copulas are BVN copulas with correlation parameters  $\rho_{12},\rho_{23},\rho_{34}$  (1st tree/level) and partial correlation parameters $\rho_{13|2}, \rho_{24|3}, \rho_{14|23}$ (2nd and 3rd tree/level),
the resulting distribution is the QVN  with mean vector $\bmu=\bigl(l(\pi_1),l(\pi_2),l(\pi_3),l(\pi_4)\bigr)^\top$ and variance covariance matrix
$\Sbf=\begin{pmatrix}
\sigma_1^2 &\rho_{12}\sigma_1\s_2 &\rho_{13}\sigma_1\s_3&\rho_{14}\sigma_1\s_4\\
\rho_{12}\sigma_1\sigma_2 & \sigma_2^2&\rho_{23}\sigma_2\s_3&\rho_{24}\sigma_2\s_4\\
\rho_{13}\sigma_1\sigma_3 &\rho_{23}\sigma_2\s_3& \sigma_3^2 &\rho_{34}\sigma_3\sigma_4\\
\rho_{14}\sigma_1\sigma_4&\rho_{24}\sigma_2\sigma_4  & \rho_{34}\sigma_3\sigma_4
&\sigma_4^2
\end{pmatrix},$ where
$\r_{13}=\r_{13|2}\sqrt{1-\r_{12}^2}\sqrt{1-\r_{23}^2} +\r_{12}\r_{23}$, $\r_{24}=\r_{24|3}\sqrt{1-\r_{23}^2}\sqrt{1-\r_{34}^2} + \r_{23}\r_{34}$,
$\r_{14}=\r_{14|2}\sqrt{1-\r_{12}^2}\sqrt{1-\r_{24}^2}+\r_{12}\r_{24}$, 
$\r_{14|2}=\r_{14|23}\sqrt{1-\r_{13|2}^2}\sqrt{1-\r_{34|2}^2}+\r_{13|2}\r_{34|2}$, 
$\r_{13|2}=(\r_{13}-\r_{12}\r_{23})/\sqrt{1-\r_{12}^2}/\sqrt{1-\r_{23}^2}$ and 
$\r_{34|2}=(\r_{34}-\r_{23}\r_{24})/\sqrt{1-\r_{23}^2}/\sqrt{1-\r_{24}^2}$.
Therefore,  the between-studies model in (\ref{copula-between-norm}) 
 assumes
that $(\X_1,\X_2)$ is QVN distributed, i.e., $
\X
\sim
\mbox{QVN}
\bigl(\bmu,\Sbf\bigr).
$

With some algebra it can be shown that the joint likelihood in (\ref{mixed-cop-likelihood})
becomes
\begin{multline*}
L(\pi_1,\pi_2,\pi_3,\pi_4,\sigma_1,\sigma_2,\sigma_3,\sigma_4,\rho_{12},\rho_{13},\rho_{14},\rho_{23},\rho_{24},\rho_{34})
=\\\prod_{i=1}^N\int_{-\infty}^{\infty}\int_{-\infty}^{\infty}\int_{-\infty}^{\infty}\int_{-\infty}^{\infty}
\prod_{j=1}^4g\Bigl(y_{ij};n_{ij},l^{-1}(x_j)\Bigr)\phi_{1234}(x_1,x_2,x_3,x_4;\bmu,\Sbf)dx_1dx_2dx_3dx_4. 
\end{multline*}
where $\phi_{1234}(\cdot;\bmu,\Sbf)$ is the QVN density with mean vector $\bmu$  and variance covariance matrix $\Sbf$. Hence,  this model is the same as the quadrivariate GLMM\cite{hoyer&kuss-2016-smmr}.

\subsubsection{Relationship with a vine copula model with beta-binomial margins}

Hoyer and Kuss  \cite{hoyer&kuss-2017-sim}  proposed a vine copula model with beta-binomial margins in this context. This model is actually an approximated likelihood method for the estimation of the D-vine copula mixed model with beta margins for the latent pair of sensitivity and specificity in (\ref{withinBinom}) and (\ref{copula-between-norm}).
They attempt to approximate the likelihood  in (\ref{mixed-cop-likelihood})  with the likelihood of a copula model for observed discrete variables which have beta-binomial margins.

The approximation that they suggest is 
$$
L(\pi_1,\pi_2,\pi_3,\pi_4,\g_1,\g_2,\g_3,\g_4,\thbf)\approx
\prod_{i=1}^Nc_{1234}\Bigl(H(y_{i1};n_{i1},\pi_1,\g_1),H(y_{i2};n_{i2},\pi_2,\g_2)
,H(y_{i3};n_{i3},\pi_3,$$

\vspace{-3ex}

\noindent $\g_3),H(y_{i4};n_{i4},\pi_4,\g_4);\thbf\Bigr)\prod_{j=1}^4 h(y_{ij};n_{ij},\pi_j,\g_j),
$
where $H(\cdot;n,\pi,\g)$ and  $h(\cdot;n,\pi,\g)$  is the cdf and the density, respectively, of the the Beta-Binomial($n,\pi,\g$) distribution.
In their approximation the authors also treat the observed variables which have beta-binomial distributions as being continuous, and model them under the theory for copula models with continuous margins.

Nikoloulopoulos \cite{Nikoloulopoulos2015b}  has  extensively studied the small-sample and theoretical efficiency  of this approximation in the bivariate case \cite{kuss-etal-2013}. It was clearly shown that this approximation leads  to substantial downward bias for the estimates  of the dependence and bias for the meta-analytic parameters for fully specified copula mixed models.  This evolves because there are serious problems on modelling assumptions under the case of heterogeneous  study sizes. If
the number of true positives and true negatives  do not have a common support over different studies,  then one cannot conclude that there is a copula.  
The copula is not common if the mixing distribution for binomials is common.  
Nevertheless, it is a stronger assumption to assume a common copula
for beta-binomial random variables with different parameters (the study size $n_{ij}$), compared
with a common copula for the random effects.  It is more natural to join the random effects with a copula. \cite{Nikoloulopoulos2016-letter-smmr}

\subsection{\label{computation}Maximum likelihood estimation and computational details}

Estimation of the model parameters $(\pi_1,\pi_2,\pi_3,\pi_4,\de_1,\de_2,\de_3,\de_4,\thbf)$   can be approached by the standard maximum likelihood (ML) method, by maximizing the logarithm of the joint likelihood in (\ref{mixed-cop-likelihood}). 
The estimated parameters can be obtained by 
using a quasi-Newton \cite{nash90} method applied to the logarithm of the joint likelihood.  
This numerical  method requires only the objective
function, i.e.,  the logarithm of the joint likelihood, while the gradients
are computed numerically and the Hessian matrix of the second
order derivatives is updated in each iteration. The standard errors (SE) of the ML estimates can be also obtained via the gradients and the Hessian computed numerically during the maximization process.

For D-vine copula mixed models of the form with joint likelihood as in (\ref{mixed-cop-likelihood}), numerical evaluation of the joint pmf can be achieved with the following steps:

\begin{enumerate}
\itemsep=0pt
\item Calculate Gauss-Legendre \cite{Stroud&Secrest1966}  quadrature points $\{u_q: q=1,\ldots,n_q\}$ 
and weights $\{w_q: q=1,\ldots,n_q\}$ in terms of standard uniform.
\item Convert from independent uniform random variables $\{u_{q_1}: q_1=1,\ldots,n_q\}$,  $\{u_{q_2}: q_2=1,\ldots,n_q\}$, $\{u_{q_3}: q_3=1,\ldots,n_q\}$, and $\{u_{q_4}: q_4=1,\ldots,n_q\}$ to   dependent uniform random variables that have a D-vine distribution $C(\cdot;\thbf)$:

\begin{algorithmic}[1]
\State Set  $v_{q_1}=u_{q_1}$
\State $v_{q_2|q_1}=C^{-1}_{2|1}(u_{q_2}|u_{q_1};\th_{12})$
\State $t_1=C_{1|2}(v_{q_1}|v_{q_2|q_1};\th_{12})$
\State $t_2=C_{3|1;2}^{-1}\left(u_{q_3}|t_1;\th_{12}),\th_{13|2}\right)$
\State $v_{q_3|q_1;q_2}=C_{3|2}^{-1}(t_2|v_{q_2|q_1};\th_{23})$
\State $t_3=C_{2|3}(v_{q_2|q_1}|v_{q_3|q_1;q_2};\th_{23})$
\State $t_4=C_{1|3;2}(t_1|t_2;\th_{13|2})$
\State $t_5=C_{4|1;2,3}(u_{q_4}|t_4;\th_{14|23})$
\State $t_6=C^{-1}_{4|2;3}(t_5|t_3;\th_{24|3})$
\State $v_{q_4|q_1;q_2,q_3}=C^{-1}_{4|3}(t_6|v_{q_3|q_1;q_2};\th_{34})$
\end{algorithmic}

The simulation algorithm of a D-vine copula in Joe \cite{joe2014} is used  to achieve this.
\item Numerically evaluate the joint pmf
$$\int_{0}^{1}\int_{0}^{1}\int_{0}^{1}\int_{0}^{1}
\prod_{j=1}^4g\Bigl(y_{ij};n_{ij},l^{-1}\bigl(F^{-1}(u_j;l(\pi_j),\g_j)\bigr)\Bigr)c_{1234}(u_1,u_2,u_3,u_4;\thbf)du_1du_2du_3du_4$$
in a quadruple sum:
\end{enumerate}

\begin{multline*}
\sum_{q_1=1}^{n_q}\sum_{q_2=1}^{n_q}\sum_{q_3=1}^{n_q}
\sum_{q_4=1}^{n_q}w_{q_1}w_{q_2}w_{q_3}w_{q_4}
g\Bigl(y_{1};n,l^{-1}\bigl(F^{-1}(v_{q_1};l(\pi_1),\g_1)\bigl)\Bigr)
g\Bigl(y_{2};n,l^{-1}\bigl(F^{-1}(v_{q_2|q_1};l(\pi_2),\\
\g_2)\bigr)\Bigr)g\Bigl(y_{3};n,l^{-1}\bigl(F^{-1}(v_{q_2q_3|q_1};l(\pi_3),\g_3)\bigr)\Bigr)g\Bigl(y_{4};n,l^{-1}\bigr(F^{-1}(v_{q_4|q_1;q_2,q_3};l(\pi_4),\g_4)\bigl)\Bigr).
\end{multline*}

With Gauss-Legendre quadrature, the same nodes and weights
are used for different functions;
this helps in yielding smooth numerical derivatives for numerical optimization via quasi-Newton.\cite{nash90} 
The  conditional copula cdfs  $C(v|u;\th)$ and their inverses $C^{-1}(v|u;\th)$ are given in Table \ref{2fam} for   the  sufficient list of parametric families of copulas  for meta-analysis of diagnostic test accuracy studies.\cite{Nikoloulopoulos2015b,Nikoloulopoulos2015c,Nikoloulopoulos-2016-SMMR,Nikoloulopoulos2018-AStA} Since  the copula parameter $\th$ of each   family has different range, in the sequel we reparametrize them via their Kendall’s $\tau$; that is comparable across families. The functional relationships between $\th$ and $\tau$ for each family are also given in Table \ref{2fam}.

\setlength{\tabcolsep}{8pt}
\begin{landscape}
\begin{table}[!h]
\begin{center}

\vspace{3.5cm} 

\caption{\label{2fam}Parametric families of bivariate copulas and their Kendall's $\tau$ as a strictly increasing function of the copula parameter $\theta$.}

\begin{tabular}{cccc}
\hline\\
Copula &$C(v|u;\th)$& $C^{-1}(v|u;\th)$& 
$\tau$\\\\\hline\\
BVN &$\Phi\Bigl(\frac{\Phi^{-1}(v)-\th\Phi^{-1}(u)}{\sqrt{1-\th^2}}\Bigr)$& $\Phi\Bigl(\sqrt{1-\th^2}\Phi^{-1}(v)+\th\Phi^{-1}(u)\Bigr)$ 
&$\frac{2}{\pi}\arcsin(\th)\quad ,\quad -1\leq\th\leq1$\\\\
Frank &$e^{-\th u}\Bigl[\frac{1-e^{-\th}}{1-e^{-\th v}}-(1-e^{-\th u})\Bigr]^{-1}$&$
-\frac{1}{\theta}\log\left[1-\frac{1-e^{-\th}}{(v^{-1}-1)e^{-\th u}+1}\right]
$
&$\begin{array}{ccc}
1-4\theta^{-1}-4\theta^{-2}\int_\theta^0\frac{t}{e^t-1}dt &,& \th<0\\
1-4\theta^{-1}+4\theta^{-2}\int^\theta_0\frac{t}{e^t-1}dt &,& \th>0\\
\end{array}$\\\\
Clayton  &$\Bigl[1+u^\th(v^{-\th}-1)\Bigr]^{-1-1/\th}$&$\Bigl[(v^{-\theta/(1+\theta)}-1)u^{-\th}+1\Bigr]^{-1/\theta}$
 &$\th/(\th+2)\quad ,\quad \th>0$\\\\
Cln90$^\circ$ &$\Bigl[1+(1-u)^\th(v^{-\th}-1)\Bigr]^{-1-1/\th}$&$\Bigl[(v^{-\theta/(1+\theta)}-1)(1-u)^{-\th}+1\Bigr]^{-1/\theta}$&
$-\th/(\th+2)\quad ,\quad \th>0$\\\\
Cln180$^\circ$ &$1-\Bigl[1+(1-u)^\th((1-v)^{-\th}-1)\Bigr]^{-1-1/\th}$&$1-\Bigl[\bigl\{(1-v)^{-\theta/(1+\theta)}-1\bigr\}(1-u)^{-\th}+1\Bigr]^{-1/\theta}$
&$\th/(\th+2)\quad ,\quad \th>0$\\\\
Cln270$^\circ$ &$1-\Bigl[1+u^\th\bigl\{(1-v)^{-\th}-1\bigr\}\Bigr]^{-1-1/\th}$&$1-
\Bigl[\bigl\{(1-v)^{-\theta/(1+\theta)}-1\bigr\}u^{-\th}+1\Bigr]^{-1/\theta}$
&$-\th/(\th+2)\quad ,\quad \th>0$\\\\
\hline
\end{tabular}

\begin{flushleft}
\begin{footnotesize}
Cln$\omega^\circ$: Clayton rotated by $\omega$ degrees. 
\end{footnotesize}  
\end{flushleft}
\end{center}
\end{table}
\end{landscape}

\section{\label{miss-section}Small-sample efficiency -- Misspecification of the random effects distribution}

An extensive simulation study is conducted  
(a) to gauge the small-sample efficiency of the ML and approximated likelihood in Hoyer and Kuss  \cite{hoyer&kuss-2017-sim} (hereafter HK) methods, and 
(b) to investigate in detail 
the  misspecification of the parametric margin or  family of copulas of the random effects distribution.    

\begin{figure}[!h]
\begin{center}
\begin{footnotesize}
\begin{tabular}{|cc|}
\hline
Diabetes& Arthritis \\\hline

\includegraphics[width=0.5\textwidth]{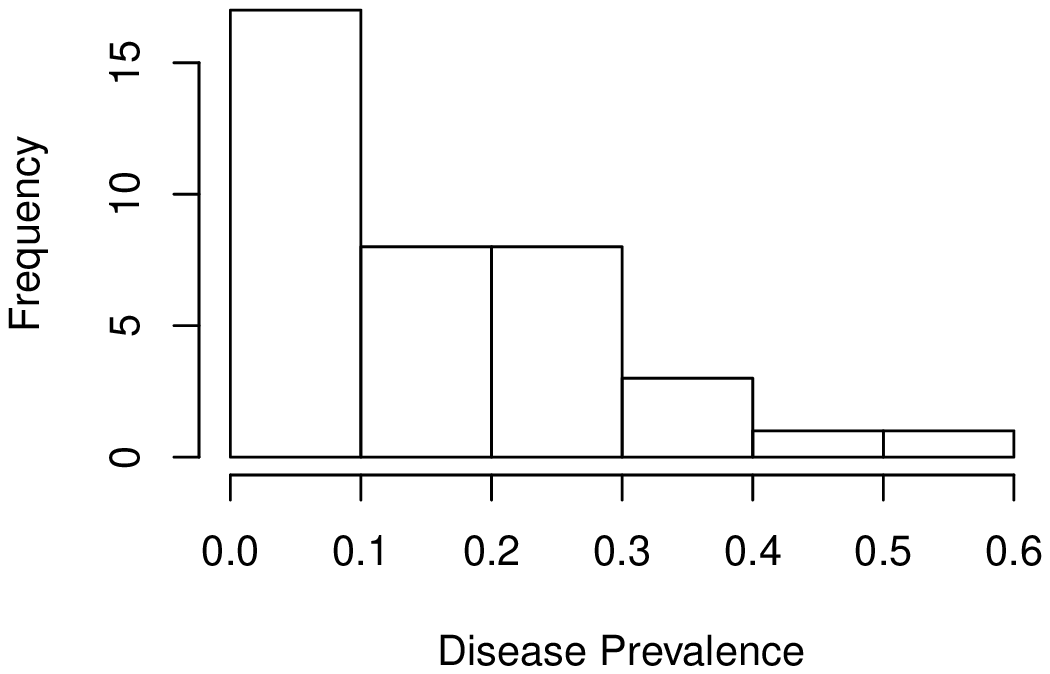}
&

\includegraphics[width=0.5\textwidth]{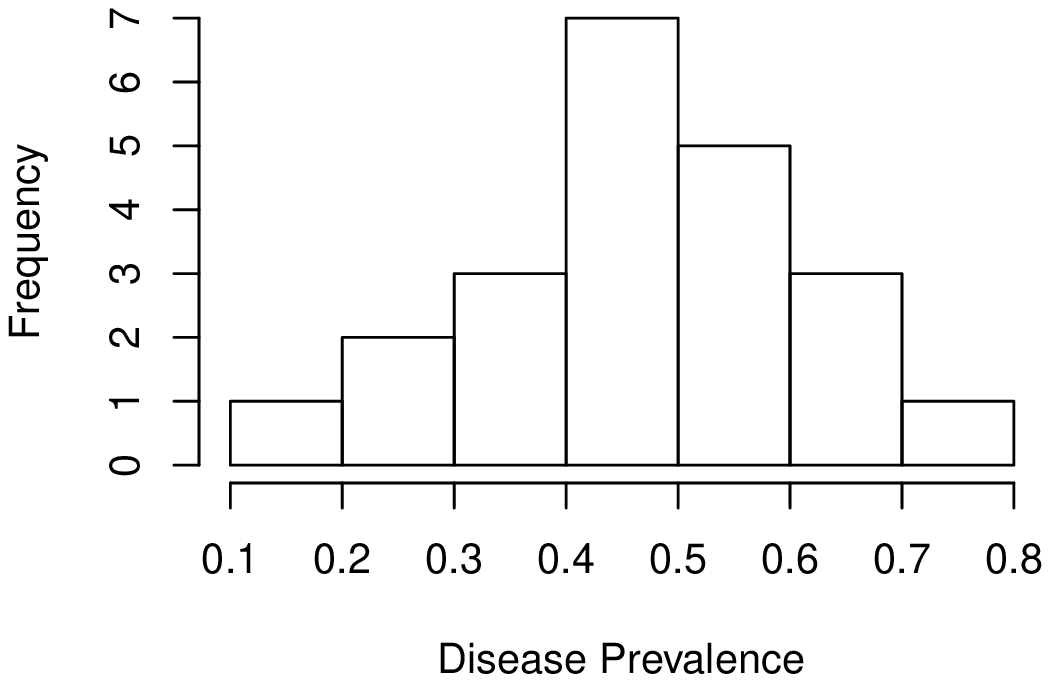}\\\hline
\end{tabular}
\end{footnotesize}
\caption{\label{prev}Disease prevalences for the arthritis and diabetes data. }
\end{center}
\vspace{-0.5cm}
\end{figure}

\begin{figure}[!h]
\begin{center}
\begin{footnotesize}
\begin{tabular}{|cc|}
\hline
Diabetes& Arthritis \\\hline

\includegraphics[width=0.5\textwidth]{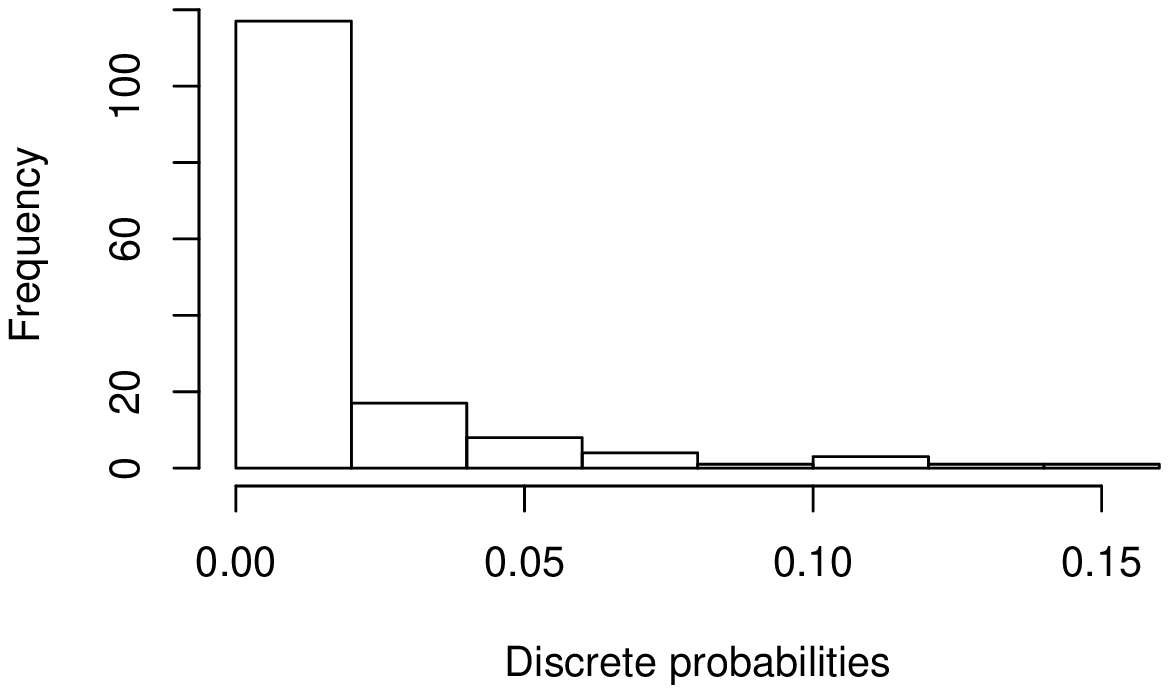}
&

\includegraphics[width=0.5\textwidth]{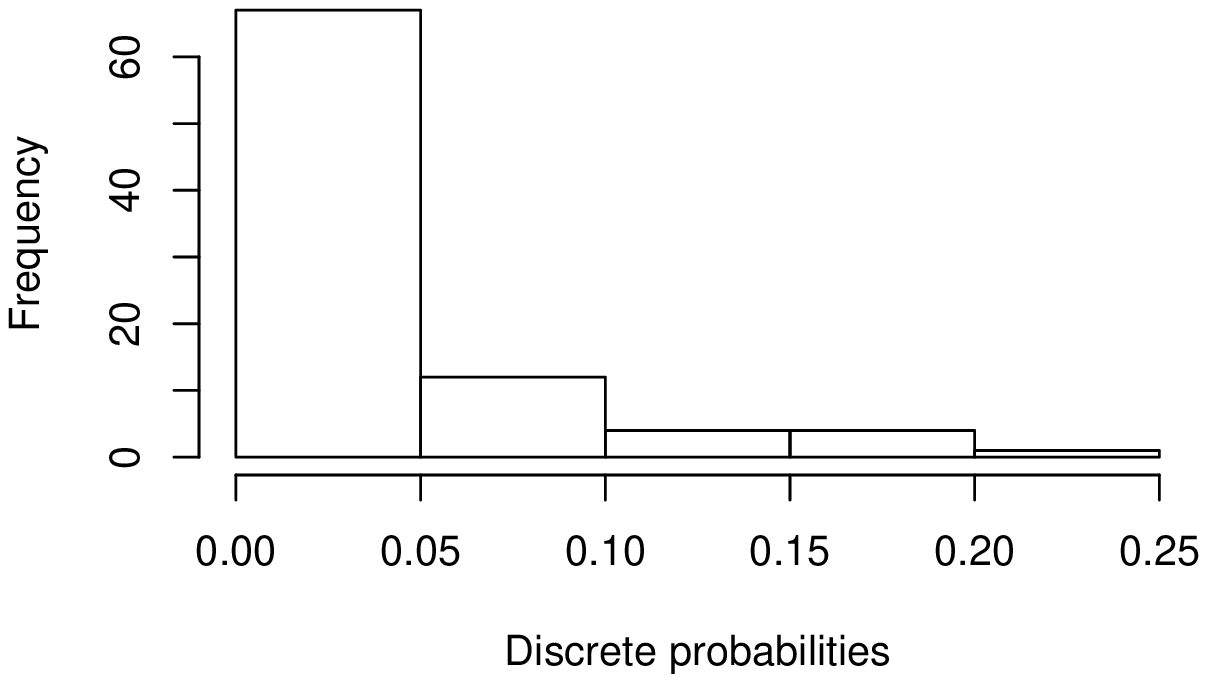}\\\hline
\end{tabular}
\end{footnotesize}
\caption{\label{prob} The size of the the discrete probabilities assuming independence for the arthritis and diabetes data.}
\end{center}
\vspace{-0.5cm}
\end{figure}

We set the sample size $N$, the study size $n$, the true univariate and Kendall's $\tau$  parameters, and the disease prevalence $\pi$ to mimic both  the arthritis \cite{Nishimura-etal2007} and the diabetes \cite{Bennett-etal2007,Kodama-etal2013} data sets. 
These example meta-analyses have different disease prevalences as depicted in Figure \ref{prev}. 
To quantify this we fit the model ignoring the dependence among the random effects, that is we assume independence. Figure \ref{prob} depicts the size of the estimated beta-binomial discrete probabilities  under the independence assumption for both datasets. As revealed the diabetes data  posses very small individual-study discrete probabilities.

More specifically, we randomly generate samples of size {\bf $N = 22,38$}  from the D-vine copula mixed model with BVN linking copulas and both  normal and beta margins. The simulation process is as below:

\vspace{-2ex}

\begin{enumerate}
\itemsep=-5pt
 \item Simulate $(u_1,u_2,u_3,u_4)$ from a D-vine distribution $C(;\taubf)$;  $\taubf$ is converted 
to the  dependence parameters $\thbf$ via the relations  in Table \ref{2fam}.  

\item Convert  to proportions via $x_j=l^{-1}\Bigl(F_j^{-1}\bigl(u_j,l(\pi_j),\de_j\bigr)\Bigr)$.  
\item Draw the number of diseased $n_{1}$ from a $B(n,\pi)$ distribution.
\item Set  $n_2=n-n_1$ and generate $y_j$ from a $B(n_j,x_j)$ for $j=1,\ldots,4$. 
\end{enumerate}

Tables \ref{norm-sim} and \ref{beta-sim} contain the resultant biases, root mean square errors (RMSEs) and standard deviations (SDs), along with average theoretical variances, for the MLEs under different copula choices and margins. The theoretical variances of the maximum likelihood estimates (MLEs) are obtained via the gradients and the Hessian that were computed numerically during the maximization process. We also provide biases, RMSEs and SDs for the HK estimates under the ‘true’ model, that is, the BVN copula mixed model with normal (Table \ref{norm-sim}) and beta (Table \ref{beta-sim})   margins.

Conclusions from the values in the 2 tables and other computations that we have done  are the following:

\vspace{-2ex}

\begin{itemize}
\itemsep=-5pt
\item ML  with  the  true copula mixed  model is highly efficient according to the simulated biases and standard deviations.

\item The MLEs of the meta-analytic parameters are slightly underestimated under copula misspecification.

\item The SDs are rather robust to the copula misspecification.

\item The meta-analytic MLEs and SDs are not robust to the margin misspecification, while 
the MLE of $\tau$ and  its SD is.

\item The HK method is not efficient for the univariate marginal parameters when there is more discretization (larger individual-study discrete probabilities).

\item The efficiency of the HK method is low for the Kendall's $\tau$ association parameters. The parameters are substantially underestimated when there is more discretization (larger individual probabilities). 

\item The HK method yields estimates that are almost as good as the MLEs for the univariate parameters when there is less discretization (small individual probabilities) and the true distribution  used for the marginal modelling of
the latent proportions is  the Beta$(\pi,\gamma)$. 
\item When there is less disretization 
the number of quadrature points  should be at least $n_q=30$ rather than $n_q=15$ which is a sufficient number of quadrature points when there are large individual probabilities. 
\end{itemize}

\setlength{\tabcolsep}{8pt}
\begin{landscape}
\begin{table}[!h]
\begin{center}

\begin{small}

 \centering
  \caption{\label{norm-sim}Small sample of sizes $N = 22$ simulations ($10^3$ replications; $n_q=15$) from the D-vine copula mixed model with BVN copulas and normal margins and resultant biases,  root mean square errors (RMSE) and standard deviations (SD), along with the square root of the average theoretical variances ($\sqrt{\bar V}$), for the MLEs  under different copula choices and margins. We also provide biases, RMSEs and SDs for the HK estimates under the true model. We set the  disease prevalence $\pi$ to mimic  the arthritis data. }
    \begin{tabular}{lllcccccccccccccc}
    \toprule
          &  margin &copula  & $\pi_1=$ & $\pi_2=$ & $\pi_3=$ & $\pi_4=$ & $\s_1=$ & $\s_2=$ & $\s_3=$ & $\s_4=$ & $\tau_{12}=$ & $\tau_{23}=$ & $\tau_{34}=$ & $\tau_{13|2}=$ & $\tau_{24|3}=$ & $\tau_{14|23}=$ \\
                    &   &  & $0.7$ & $0.8$ & $0.7$ & $0.95$ & $0.7$ & $1$ & $0.7$ & $0.8$ & $-0.3$ & $0.1$ & $-0.4$ & $0.5$ & $0.4$ & $0.2$ \\

    \midrule
 Bias        & $^\dag$ normal & BVN   & -0.003 & -0.004 & -0.003 & -0.001 & -0.017 & -0.013 & -0.023 & -0.033 & -0.001 & 0.009 & -0.008 & 0.038 & 0.058 & 0.116 \\
          & beta  &    & -0.018 & -0.042 & -0.018 & -0.015 &  &  &  &  & -0.001 & 0.003 & -0.003 & 0.033 & 0.040 & 0.095 \\
          & normal & Frank   & -0.001 & -0.005 & -0.002 & -0.001 & -0.016 & 0.001 & -0.014 & -0.031 & 0.002 & 0.023 & -0.013 & 0.042 & 0.064 & 0.123 \\
          & beta  &   & -0.016 & -0.041 & -0.017 & -0.015 &  & &  & & 0.001 & 0.017 & -0.015 & 0.039 & 0.046 & 0.092 \\
          & normal & Cln90$^\circ$ & -0.004 & -0.001 & -0.002 & 0.000 & -0.012 & 0.011 & -0.020 & -0.022 & 0.050 & 0.066 & 0.071 & 0.006 & 0.066 & 0.083 \\
          & beta  & & -0.019 & -0.041 & -0.017 & -0.015 & &  &  &  & 0.053 & 0.076 & 0.076 & 0.006 & 0.067 & 0.082 \\
          & normal & Cln270$^\circ$ & -0.001 & -0.007 & -0.004 & -0.001 & -0.011 & 0.044 & -0.005 & -0.026 & 0.057 & 0.065 & 0.019 & 0.016 & 0.074 & 0.076 \\
          & beta  &  & -0.018 & -0.047 & -0.020 & -0.015 &  & &  &  & 0.062 & 0.050 & 0.027 & 0.005 & 0.038 & 0.032 \\
          & $^\ddag$ beta  & BVN   & -0.010 & -0.048 & -0.012 & -0.014 &  &  &  & & 0.081 & 0.020 & 0.249 & -0.124 & -0.244 & -0.178 \\
\hline
  SD        & $^\dag$ normal & BVN   & 0.036 & 0.037 & 0.035 & 0.010 & 0.145 & 0.197 & 0.134 & 0.173 & 0.166 & 0.168 & 0.173 & 0.162 & 0.206 & 0.447 \\
          & beta  &   & 0.034 & 0.038 & 0.033 & 0.012 & 0.030 & 0.042 & 0.028 & 0.017 & 0.163 & 0.166 & 0.168 & 0.161 & 0.204 & 0.424 \\
          & normal & Frank   & 0.036 & 0.038 & 0.036 & 0.010 & 0.146 & 0.202 & 0.139 & 0.177 & 0.173 & 0.170 & 0.189 & 0.162 & 0.213 & 0.465 \\
          & beta  &  & 0.035 & 0.038 & 0.034 & 0.013 & 0.030 & 0.043 & 0.029 & 0.017 & 0.168 & 0.167 & 0.185 & 0.164 & 0.208 & 0.446 \\
          & normal & Cln90$^\circ$ & 0.036 & 0.039 & 0.036 & 0.010 & 0.148 & 0.219 & 0.147 & 0.182 & 0.167 & 0.174 & 0.204 & 0.180 & 0.219 & 0.468 \\
          & beta  &  & 0.034 & 0.038 & 0.034 & 0.013 & 0.029 & 0.047 & 0.030 & 0.019 & 0.162 & 0.164 & 0.204 & 0.178 & 0.217 & 0.442 \\
          & normal & Cln270$^\circ$ & 0.036 & 0.040 & 0.035 & 0.010 & 0.146 & 0.228 & 0.154 & 0.182 & 0.166 & 0.184 & 0.234 & 0.177 & 0.222 & 0.479 \\
          & beta  &  & 0.035 & 0.039 & 0.034 & 0.013 & 0.032 & 0.048 & 0.034 & 0.017 & 0.175 & 0.192 & 0.228 & 0.179 & 0.216 & 0.452 \\
          & $^\ddag$ beta  & BVN   & 0.033 & 0.038 & 0.032 & 0.012 & 0.024 & 0.037 & 0.024 & 0.015 & 0.132 & 0.128 & 0.114 & 0.128 & 0.121 & 0.109 \\
\hline
   $\sqrt{\bar V}$       & $^\dag$ normal & BVN   & 0.027 & 0.030 & 0.031 & 0.009 & 0.101 & 0.148 & 0.117 & 0.166 & 0.114 & 0.138 & 0.155 & 0.138 & 0.237 & 0.571 \\
          & beta  &    & 0.025 & 0.028 & 0.028 & 0.011 & 0.021 & 0.030 & 0.025 & 0.014 & 0.120 & 0.138 & 0.153 & 0.131 & 0.159 & 0.678 \\
          & normal & Frank   & 0.027 & 0.029 & 0.030 & 0.009 & 0.070 & 0.149 & 0.113 & 0.165 & 0.112 & 0.136 & 0.162 & 0.135 & 0.165 & 0.729 \\
          & beta  &    & 0.025 & 0.027 & 0.028 & 0.011 & 0.021 & 0.030 & 0.026 & 0.015 & 0.132 & 0.140 & 0.161 & 0.127 & 0.194 & 0.452 \\
          & normal & Cln90$^\circ$ & 0.027 & 0.030 & 0.032 & 0.009 & 0.099 & 0.155 & 0.120 & 0.167 & 0.094 & 0.144 & 0.159 & 0.137 & 0.174 & 0.625 \\
          & beta  &  & 0.025 & 0.027 & 0.029 & 0.011 & 0.020 & 0.031 & 0.025 & 0.015 & 0.097 & 0.136 & 0.157 & 0.136 & 0.174 & 0.584 \\
          & normal & Cln270$^\circ$ & 0.027 & 0.028 & 0.031 & 0.009 & 0.101 & 0.142 & 0.123 & 0.167 & 0.099 & 0.141 & 0.183 & 0.138 & 0.205 & 0.479 \\
          & beta  &  & 0.026 & 0.024 & 0.029 & 0.010 & 0.021 & 0.027 & 0.027 & 0.014 & 0.104 & 0.143 & 0.188 & 0.137 & 0.067 & 0.319 \\
          & $^\ddag$ beta  & BVN   & 0.030 & 0.034 & 0.030 & 0.011 & 0.024 & 0.033 & 0.023 & 0.013 & 0.113 & 0.117 & 0.080 & 0.104 & 0.082 & 0.091 \\
\hline
 RMSE          & $^\dag$ normal & BVN   & 0.036 & 0.037 & 0.035 & 0.010 & 0.146 & 0.197 & 0.136 & 0.176 & 0.166 & 0.168 & 0.173 & 0.166 & 0.214 & 0.461 \\
          & beta  &    & 0.038 & 0.056 & 0.038 & 0.019 &  &  & &  & 0.163 & 0.166 & 0.168 & 0.164 & 0.208 & 0.434 \\
          & normal & Frank   & 0.036 & 0.039 & 0.036 & 0.010 & 0.147 & 0.202 & 0.140 & 0.180 & 0.173 & 0.172 & 0.189 & 0.168 & 0.223 & 0.481 \\
          & beta  &   & 0.038 & 0.056 & 0.038 & 0.020 &  &  &  &  & 0.168 & 0.168 & 0.185 & 0.168 & 0.213 & 0.455 \\
          & normal & Cln90$^\circ$ & 0.036 & 0.039 & 0.036 & 0.010 & 0.149 & 0.220 & 0.149 & 0.183 & 0.174 & 0.187 & 0.216 & 0.180 & 0.228 & 0.475 \\
          & beta  & & 0.039 & 0.056 & 0.038 & 0.020 &  &  &  &  & 0.171 & 0.181 & 0.217 & 0.178 & 0.227 & 0.450 \\
          & normal & Cln270$^\circ$ & 0.036 & 0.041 & 0.036 & 0.010 & 0.147 & 0.232 & 0.154 & 0.184 & 0.176 & 0.196 & 0.235 & 0.178 & 0.234 & 0.485 \\
          & beta  &  & 0.039 & 0.060 & 0.039 & 0.020 &  & &  &  & 0.186 & 0.199 & 0.230 & 0.179 & 0.220 & 0.453 \\
          & $^\ddag$ beta  & BVN   & 0.034 & 0.061 & 0.034 & 0.018 &  &  &  &  & 0.154 & 0.129 & 0.274 & 0.178 & 0.273 & 0.208 \\
    \bottomrule
    \end{tabular}%
  \label{tab:addlabel}%
 \end{small}
 \vspace{-1ex} 
\begin{flushleft}
\begin{footnotesize}
Cln$\omega^\circ$: Clayton rotated by $\omega$ degrees; $^\dag$: The resulting model is the same as the GLMM; 
$^\ddag$: HK approximated likelihood method.
\end{footnotesize}  
\end{flushleft}
\end{center}
\end{table}
\end{landscape}

\setlength{\tabcolsep}{7pt}
\begin{landscape}
\begin{table}[!h]
\begin{center}

\begin{small}

  \centering
  \caption{\label{beta-sim}Small sample of sizes $N = 22$ simulations ($10^3$ replications; ; $n_q=15$) from the D-vine copula mixed model with BVN copulas and beta margins and resultant biases,  root mean square errors (RMSE) and standard deviations (SD), along with the square root of the average theoretical variances ($\sqrt{\bar V}$), for the MLEs  under different copula choices and margins. We also provide biases, RMSEs and SDs for the HK estimates under the true model. We set the  disease prevalence $\pi$ to mimic  the arthritis data. }
    \begin{tabular}{lllcccccccccccccc}
    \toprule
          & margin  & copula & $\pi_1=$ & $\pi_2=$ & $\pi_3=$ & $\pi_4=$ & $\g_1=$ & $\g_2=$ & $\g_3=$ & $\g_4=$ & $\tau_{12}=$ & $\tau_{23}=$ & $\tau_{34}=$ & $\tau_{13|2}=$ & $\tau_{24|3}=$ & $\tau_{14|23}=$ \\
       &   &  & $0.7$ & $0.8$ & $0.7$ & $0.95$ & $0.1$ & $0.15$ & $0.1$ & $0.02$ & $-0.3$ & $0.1$ & $\tau_{34}=-0.4$ & $0.5$ & $0.4$ & $0.2$ \\
\hline
Bias          & $^\dag$ normal & BVN   & 0.020 & 0.044 & 0.019 & 0.008 &  &  & &  & 0.000 & 0.008 & -0.006 & 0.042 & 0.070 & 0.135 \\
          & beta  &   & 0.000 & -0.001 & 0.000 & 0.000 & -0.004 & -0.006 & -0.004 & -0.001 & -0.003 & 0.002 & -0.015 & 0.038 & 0.052 & 0.110 \\
          & normal & Frank   & 0.019 & 0.041 & 0.018 & 0.007 &  &  &  &  & 0.002 & 0.020 & -0.016 & 0.043 & 0.082 & 0.137 \\
          & beta  &    & 0.000 & -0.002 & 0.000 & -0.001 & -0.004 & -0.003 & -0.002 & -0.001 & -0.002 & 0.017 & -0.025 & 0.041 & 0.063 & 0.109 \\
          & normal & Cln90$^\circ$ & 0.019 & 0.047 & 0.019 & 0.008 &  &  &  &  & 0.044 & 0.043 & 0.057 & 0.000 & 0.069 & 0.061 \\
          & beta  &  & -0.002 & 0.000 & 0.000 & 0.000 & -0.004 & 0.000 & -0.002 & 0.000 & 0.054 & 0.062 & 0.043 & 0.007 & 0.071 & 0.060 \\
          & normal & Cln270$^\circ$ & 0.021 & 0.044 & 0.018 & 0.007 &  &  & && 0.042 & 0.070 & 0.015 & 0.025 & 0.099 & 0.101 \\
          & beta  &  & 0.000 & -0.003 & 0.000 & 0.000 & -0.001 & 0.006 & 0.000 & 0.000 & 0.049 & 0.050 & 0.019 & 0.016 & 0.071 & 0.064 \\
          & $^\ddag$ beta  & BVN   & 0.012 & -0.008 & 0.006 & 0.000 & -0.019 & -0.020 & -0.019 & -0.003 & 0.124 & -0.005 & 0.270 & -0.134 & -0.265 & -0.177 \\\hline
      SD    & $^\dag$ normal & BVN  & 0.038 & 0.037 & 0.037 & 0.008 & 0.177 & 0.265 & 0.154 & 0.161 & 0.165 & 0.176 & 0.186 & 0.154 & 0.247 & 0.486 \\
          & beta  &    & 0.035 & 0.035 & 0.034 & 0.008 & 0.033 & 0.045 & 0.030 & 0.009 & 0.162 & 0.167 & 0.183 & 0.153 & 0.240 & 0.477 \\
          & normal & Frank   & 0.038 & 0.038 & 0.038 & 0.008 & 0.177 & 0.269 & 0.157 & 0.164 & 0.173 & 0.176 & 0.203 & 0.154 & 0.260 & 0.501 \\
          & beta  &    & 0.035 & 0.036 & 0.036 & 0.009 & 0.033 & 0.048 & 0.032 & 0.010 & 0.168 & 0.166 & 0.199 & 0.153 & 0.252 & 0.485 \\
          & normal & Cln90$^\circ$ & 0.038 & 0.038 & 0.037 & 0.008 & 0.185 & 0.277 & 0.172 & 0.169 & 0.169 & 0.190 & 0.229 & 0.178 & 0.261 & 0.514 \\
          & beta  &  & 0.035 & 0.036 & 0.035 & 0.008 & 0.033 & 0.049 & 0.035 & 0.010 & 0.164 & 0.173 & 0.229 & 0.173 & 0.258 & 0.509 \\
          & normal & Cln270$^\circ$ & 0.038 & 0.039 & 0.038 & 0.008 & 0.176 & 0.298 & 0.168 & 0.169 & 0.160 & 0.180 & 0.249 & 0.171 & 0.267 & 0.512 \\
          & beta  &  & 0.035 & 0.037 & 0.035 & 0.008 & 0.035 & 0.052 & 0.036 & 0.010 & 0.166 & 0.180 & 0.242 & 0.172 & 0.259 & 0.505 \\
          & $^\ddag$ beta  & BVN   & 0.034 & 0.036 & 0.034 & 0.008 & 0.026 & 0.040 & 0.027 & 0.009 & 0.117 & 0.113 & 0.104 & 0.132 & 0.116 & 0.105 \\
\hline
   $\sqrt{\bar V}$        &$^\dag$ normal & BVN   & 0.026 & 0.026 & 0.038 & 0.007 & 0.109 & 0.181 & 0.177 & 0.170 & 0.101 & 0.141 & 0.177 & 0.129 & 0.252 & 0.781 \\
          & beta  &    & 0.024 & 0.025 & 0.029 & 0.008 & 0.022 & 0.033 & 0.028 & 0.009 & 0.110 & 0.134 & 0.168 & 0.128 & 0.245 & 1.056 \\
          & normal & Frank   & 0.026 & 0.025 & 0.031 & 0.007 & 0.109 & 0.181 & 0.128 & 0.161 & 0.097 & 0.133 & 0.177 & 0.131 & 0.223 & 0.845 \\
          & beta  &    & 0.024 & 0.025 & 0.029 & 0.008 & 0.022 & 0.034 & 0.028 & 0.009 & 0.111 & 0.133 & 0.175 & 0.128 & 0.245 & 0.597 \\
          & normal & Cln90$^\circ$ & 0.026 & 0.027 & 0.032 & 0.007 & 0.110 & 0.191 & 0.133 & 0.161 & 0.083 & 0.159 & 0.196 & 0.133 & 0.297 & 0.772 \\
          & beta  &  & 0.024 & 0.025 & 0.030 & 0.008 & 0.022 & 0.035 & 0.028 & 0.010 & 0.092 & 0.137 & 0.156 & 0.129 & 0.296 & 0.724 \\
          & normal & Cln270$^\circ$ & 0.026 & 0.024 & 0.031 & 0.007 & 0.108 & 0.173 & 0.134 & 0.163 & 0.085 & 0.133 & 0.196 & 0.128 & 0.310 & 0.825 \\
          & beta  &  & 0.024 & 0.023 & 0.030 & 0.008 & 0.023 & 0.032 & 0.030 & 0.010 & 0.097 & 0.142 & 0.185 & 0.133 & 0.227 & 0.826 \\
          & $^\ddag$ beta  & BVN   & 0.031 & 0.033 & 0.031 & 0.008 & 0.025 & 0.037 & 0.025 & 0.009 & 0.095 & 0.097 & 0.074 & 0.102 & 0.070 & 0.085 \\
 \hline
   RMSE       &$^\dag$  normal & BVN   & 0.043 & 0.058 & 0.041 & 0.011 &  &  &  &  & 0.165 & 0.176 & 0.187 & 0.160 & 0.257 & 0.505 \\
          & beta  &   & 0.035 & 0.035 & 0.034 & 0.008 & 0.034 & 0.046 & 0.031 & 0.009 & 0.162 & 0.167 & 0.184 & 0.157 & 0.246 & 0.490 \\
          & normal & Frank   & 0.042 & 0.056 & 0.042 & 0.011 &  &  &  &  & 0.173 & 0.178 & 0.203 & 0.160 & 0.273 & 0.519 \\
          & beta  &    & 0.035 & 0.036 & 0.036 & 0.009 & 0.033 & 0.048 & 0.032 & 0.010 & 0.168 & 0.167 & 0.201 & 0.159 & 0.259 & 0.497 \\
          & normal & Cln90$^\circ$ & 0.042 & 0.060 & 0.042 & 0.011 &  &  &  &  & 0.175 & 0.195 & 0.236 & 0.178 & 0.270 & 0.517 \\
          & beta  &  & 0.035 & 0.036 & 0.035 & 0.008 & 0.033 & 0.049 & 0.035 & 0.010 & 0.173 & 0.183 & 0.233 & 0.173 & 0.268 & 0.513 \\
          & normal & Cln270$^\circ$ & 0.043 & 0.059 & 0.042 & 0.011 &  &  & && 0.165 & 0.193 & 0.250 & 0.173 & 0.285 & 0.522 \\
          & beta  &  & 0.035 & 0.037 & 0.035 & 0.008 & 0.035 & 0.053 & 0.036 & 0.010 & 0.173 & 0.187 & 0.243 & 0.173 & 0.268 & 0.509 \\
          &$^\ddag$ beta  & BVN   & 0.036 & 0.036 & 0.035 & 0.008 & 0.032 & 0.044 & 0.033 & 0.009 & 0.171 & 0.113 & 0.289 & 0.188 & 0.289 & 0.205 \\
    \bottomrule
    \end{tabular}%
  \label{tab:addlabel}%
  \end{small}
\vspace{-1ex} 
\begin{flushleft}
\begin{footnotesize}
Cln$\omega^\circ$: Clayton rotated by $\omega$ degrees; 
$^\dag$: The resulting model is the same as the GLMM; 
$^\ddag$: HK approximated likelihood method. 
\end{footnotesize}  
\end{flushleft}
\end{center}
\end{table}
\end{landscape}

These results are in line with our previous studies \cite{Nikoloulopoulos2015b,Nikoloulopoulos2015c,Nikoloulopoulos-2016-SMMR,Nikoloulopoulos2018-AStA} and also reveal that the size of the discrete probabilities is crucial for the efficiency of the HK method. For small  prevalence, the probability of any given event will decrease, and as a result, one would be  approximating `smaller steps' in the cdf, hence the HK method at this case appears to be reliable as long as the true margin is beta.

\section{\label{app-sec}  Illustrations}
We illustrate the use of the vine copula mixed model for the meta-analysis and for comparison of diagnostic accuracy studies by  using the data of three published meta-analyses \cite{Nishimura-etal2007,Bennett-etal2007,Kodama-etal2013}. These data have been previously meta-analysed in the quadrivariate case.  \cite{hoyer&kuss-2016-smmr,hoyer&kuss-2017-sim,dimou-etal2016}

We fit the D-vine copula mixed model for all  choices of parametric families of copulas for the bivariate margins of interest  and margins. For ease of interpretation, we do not mix   margins for a single test; hence,  we allow two different  margins, one for test 1 and one for test 2. For the bivariate copula $C_{23}(\cdot)$ that links the data between the two tests and  the copulas at the second and third tree we use the  BVN copulas as on the one hand they provide a wide range of dependence to account for the conditional dependence and on the other hand allow for intermediate tail dependence \cite{Hua-joe-11}. We summarize the  model in terms of largest likelihood along with the quadrivariate GLMM.  We also 
perform a separate bivariate meta-analysis for each test fitting the bivariate copula mixed model \cite{Nikoloulopoulos2015b}   to synthesize information assuming independence between the tests. Once again we summarize the best fit in terms of the likelihood principle.

Finally, we demonstrate summary receiver operating characteristic (SROC) curves and summary operating points (a pair of average sensitivity and specificity) with a confidence region and a predictive region for each test. Hence,  we provide a direct and visual  comparison between the two competing diagnostic tests.

In the first subsection we apply the ML and HK methods  to the rheumatoid arthritis data \cite{Nishimura-etal2007}, also analysed in Dimou et al. \cite{dimou-etal2016} using all the available studies and imputation to fill in the missing data. In this  dataset, there are large individual probabilities, i.e., more discretization. In the second subsection the methods are applied to the diabetes data  \cite{Bennett-etal2007,Kodama-etal2013} for which there apparently exist small individual probabilities. As emphasized in the preceding section, the size of the discrete probabilities can substantially influence the efficiency of the HK approximate likelihood method. 

\vspace{-2ex}

\subsection{Rheumatoid arthritis}
The methods are applied to the data obtained from a meta-analysis that aimed to determine whether anti-cyclic citrullinated peptide (anti-CCP) antibody  identifies more accurately patients with rheumatoid arthritis than rheumatoid factor (RF) does.\cite{dimou-etal2016}  We include $N=22$  studies that assessed both RF and anti-CCP2 antibody for diagnosing rheumatoid arthritis.

The log-likelihood showed that a D-vine copula mixed model with Clayton rotated by 270 degrees and Frank copulas both with normal margins to join the (latent) sensitivity and specificity of the first and second test, respectively,   provides the best fit (Table \ref{arthritis-res}). 

\setlength{\tabcolsep}{5pt}

\begin{table}[!h]
  \centering

  \caption{\label{arthritis-res}Maximized ML and HK log-likelihoods, estimates and standard errors (SE) of the D-vine copula mixed models for the arthritis data.}
    \begin{tabular}{c|cc|cc|cc|c|cc}
    \toprule
    test & $^\P$ copula & margin& $^\S$ copula &  margin&$^\dag$ copula & margin&test & $^\ddag$ copula & margin\\
\hline RF  & Cln270$^\circ$ & normal&  Cln270$^\circ$ & normal&  BVN & normal&RF& Cln270$^\circ$ & beta\\
 
 Anti-CCP2 & Frank & normal& Frank & normal& BVN & normal&Anti-CCP2&Frank&beta\\

    \midrule
    Param. & Estimate  & SE    & Estimate  & SE    & Estimate  & SE    & Param. & Estimate  & SE \\\hline
    $\pi_1$ & 0.679 & 0.064 & 0.683 & 0.039 & 0.674 & 0.030 & $\pi_1$ & 0.660 & 0.029 \\
    $\pi_2$ & 0.826 & 0.027 & 0.824 & 0.026 & 0.821 & 0.025 & $\pi_2$ & 0.772 & 0.034 \\
    $\pi_3$ & 0.680 & 0.056 & 0.681 & 0.026 & 0.677 & 0.032 & $\pi_3$ & 0.669 & 0.030 \\
    $\pi_4$ & 0.959 & 0.008 & 0.959 & 0.008 & 0.959 & 0.008 & $\pi_4$ & 0.949 & 0.009 \\
    $\s_1$ & 0.711 & 0.127 & 0.715 & 0.115 & 0.705 & 0.093 & $\g_1$ & 0.064 & 0.021 \\
    $\s_2$ & 1.033 & 0.197 & 1.037 & 0.193 & 1.054 & 0.176 & $\g_2$ & 0.120 & 0.034 \\
    $\s_3$ & 0.698 & 0.129 & 0.653 & 0.073 & 0.708 & 0.113 & $\g_3$ & 0.075 & 0.024 \\
    $\s_4$ & 0.784 & 0.174 & 0.815 & 0.183 & 0.791 & 0.179 & $\g_4$ & 0.025 & 0.012 \\
    $\tau_{12}$ & -0.156 & 0.113 & -0.162 & 0.135 & -0.126 & 0.134 & $\tau_{12}$ & -0.156 & 0.112 \\
    $\tau_{23}$ & -0.115 & 0.145 & -     & -     & -0.077 & 0.148 & $\tau_{23}$ & 0.051 & 0.083 \\
    $\tau_{34}$ & -0.243 & 0.185 & -0.255 & 0.185 & -0.186 & 0.176 & $\tau_{34}$ & -0.143 & 0.132 \\
    $\tau_{13|2}$ & 0.597 & 0.095 & -     & -     & 0.597 & 0.096 & $\tau_{13|2}$ & 0.308 & 0.077 \\
    $\tau_{24|3}$ & 0.331 & 0.178 & -     & -     & 0.357 & 0.174 & $\tau_{24|3}$ & 0.117 & 0.065 \\
    $\tau_{14|23}$ & 0.127 & 0.223 & -     & -     & 0.111 & 0.224 & $\tau_{14|23}$ & 0.088 & 0.070 \\\hline
    $-\log L$ & \multicolumn{2}{c}{319.4} & \multicolumn{2}{c}{329.5} & \multicolumn{2}{c}{320.3} & $-\ell$ & \multicolumn{2}{c}{322.6} \\
    \bottomrule
    \end{tabular}%
 \vspace{-1ex} 
\begin{flushleft}
\begin{footnotesize}
Cln$270^\circ$: Clayton rotated by 270 degrees;
$\P$: Best fit; $^\S$: Separate bivariate meta-analysis for each test;
$^\dag$: The resulting model is the same as the quadrivariate  GLMM; 
$^\ddag$: HK approximated likelihood method. 
\end{footnotesize}  
\end{flushleft}
\end{table}

\begin{figure}[!h]
\begin{center}
\begin{footnotesize}
\begin{tabular}{|cc|}
\hline
Rheumatoid Factor & Anti-CCP2 antibody\\\hline

\includegraphics[width=0.5\textwidth]{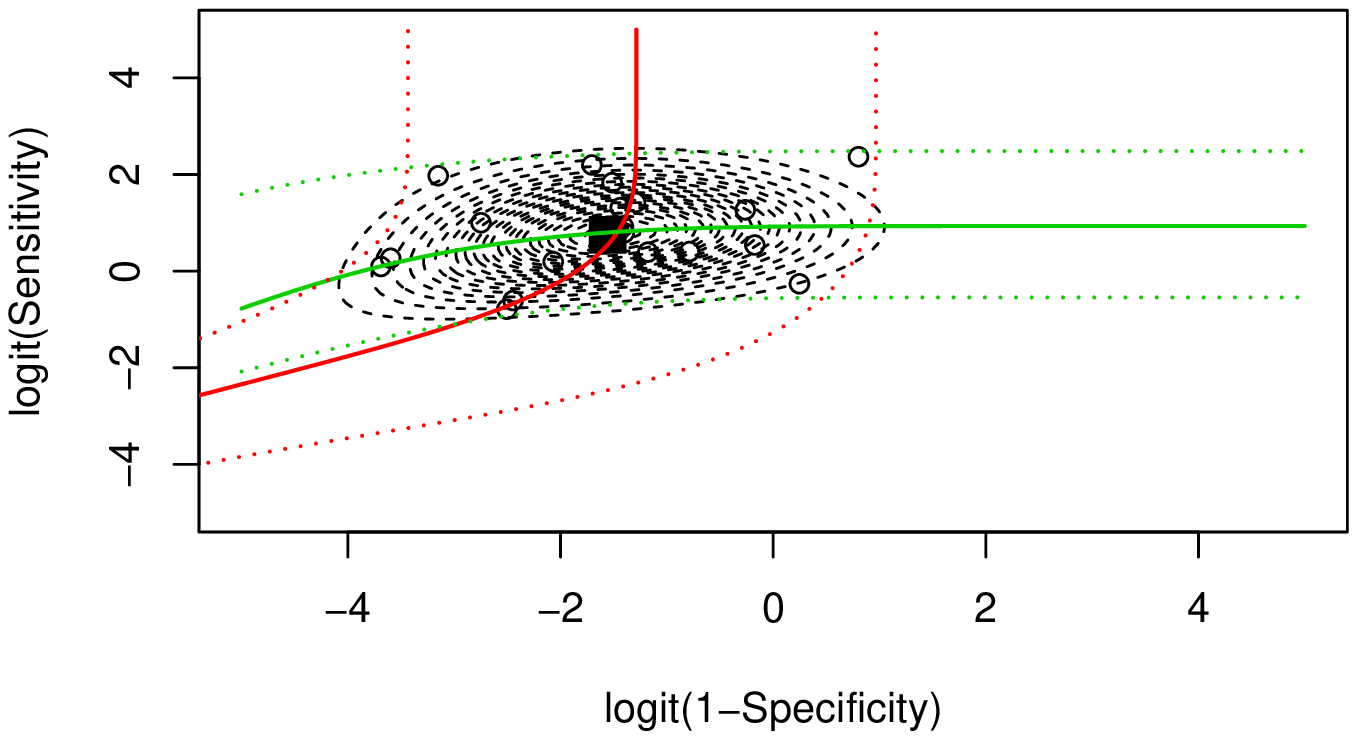}
&

\includegraphics[width=0.5\textwidth]{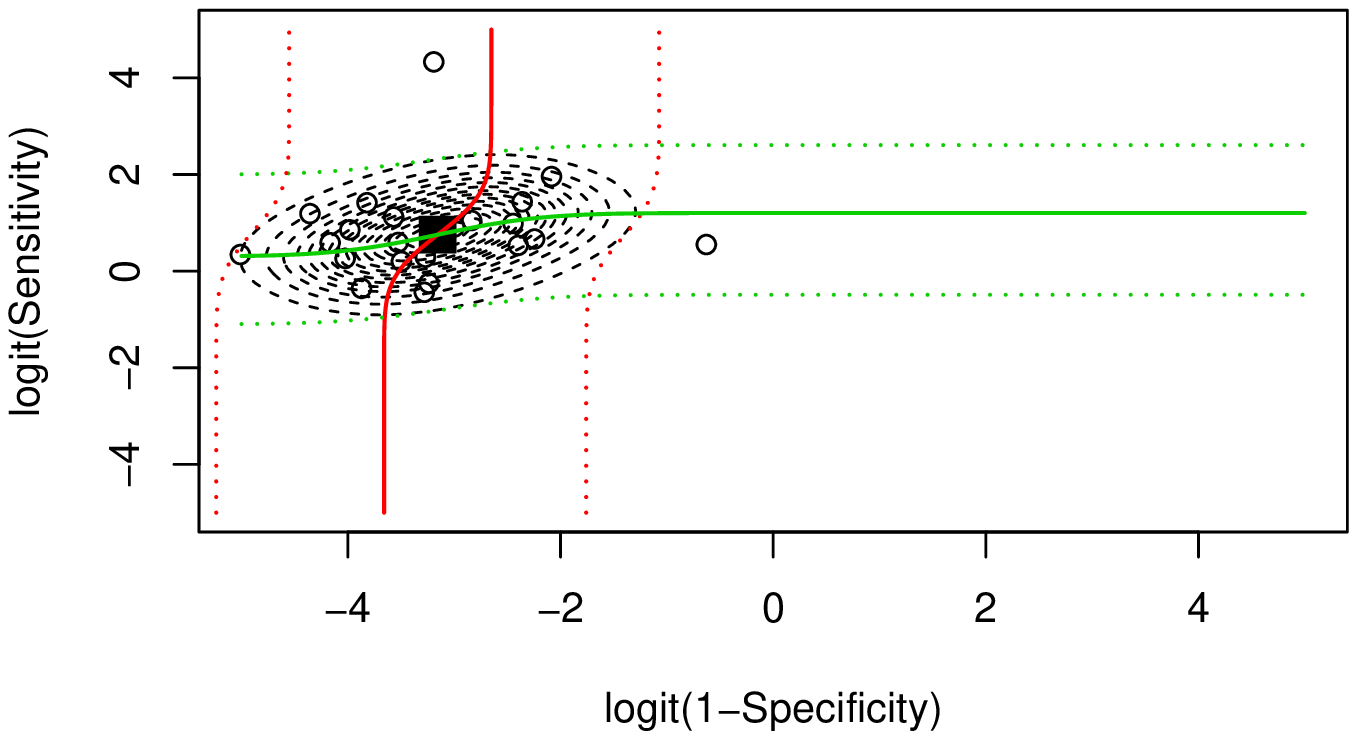}\\\hline
\end{tabular}
\end{footnotesize}
\caption{\label{ROC-arthritis}Contour plots (predictive region)  and quantile  regression curves  from the best fitted D-vine copula mixed model for the  arthritis data. Red and green lines represent the quantile  regression curves $x_1:=\widetilde{x}_1(x_2,q)$ and $x_2:=\widetilde{x}_2(x_1,q)$, respectively; for $q=0.5$ solid lines and for $q\in\{0.01,0.99\}$ dotted lines (confidence region).  median regression curve for each model. The axes are in  logit scale since  we also plot  the estimated contour plot of the random effects distribution as predictive region; this has been estimated for the logit pair of (Sensitivity, Specificity).}
\end{center}
\vspace{-1cm}
\end{figure}

\begin{figure}[!h]
\begin{center}

\begin{tabular}{|cc|}
\hline
$x_1:=\widetilde{x}_1(x_2,q=0.5)$ & $x_2:=\widetilde{x}_2(x_1,q=0.5)$\\\hline

\includegraphics[width=0.5\textwidth]{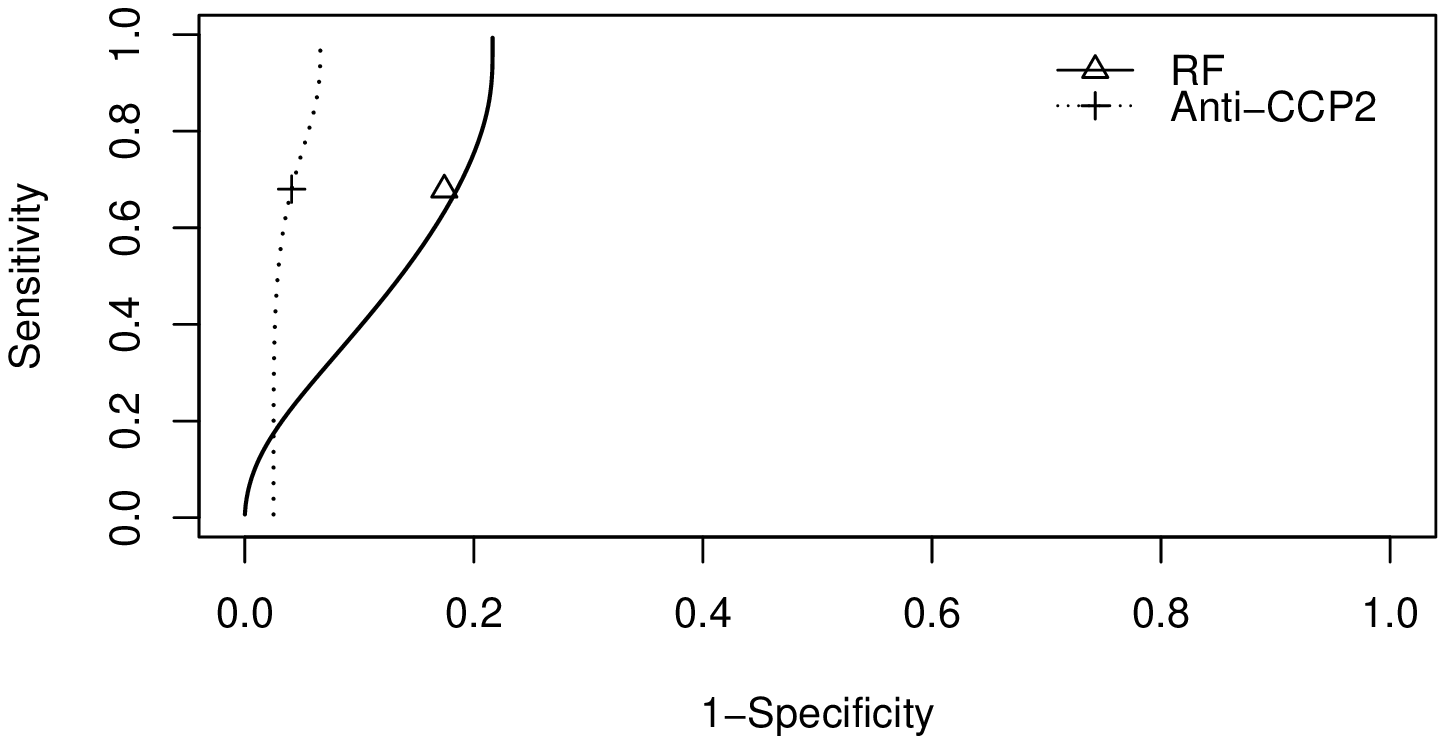}
&

\includegraphics[width=0.5\textwidth]{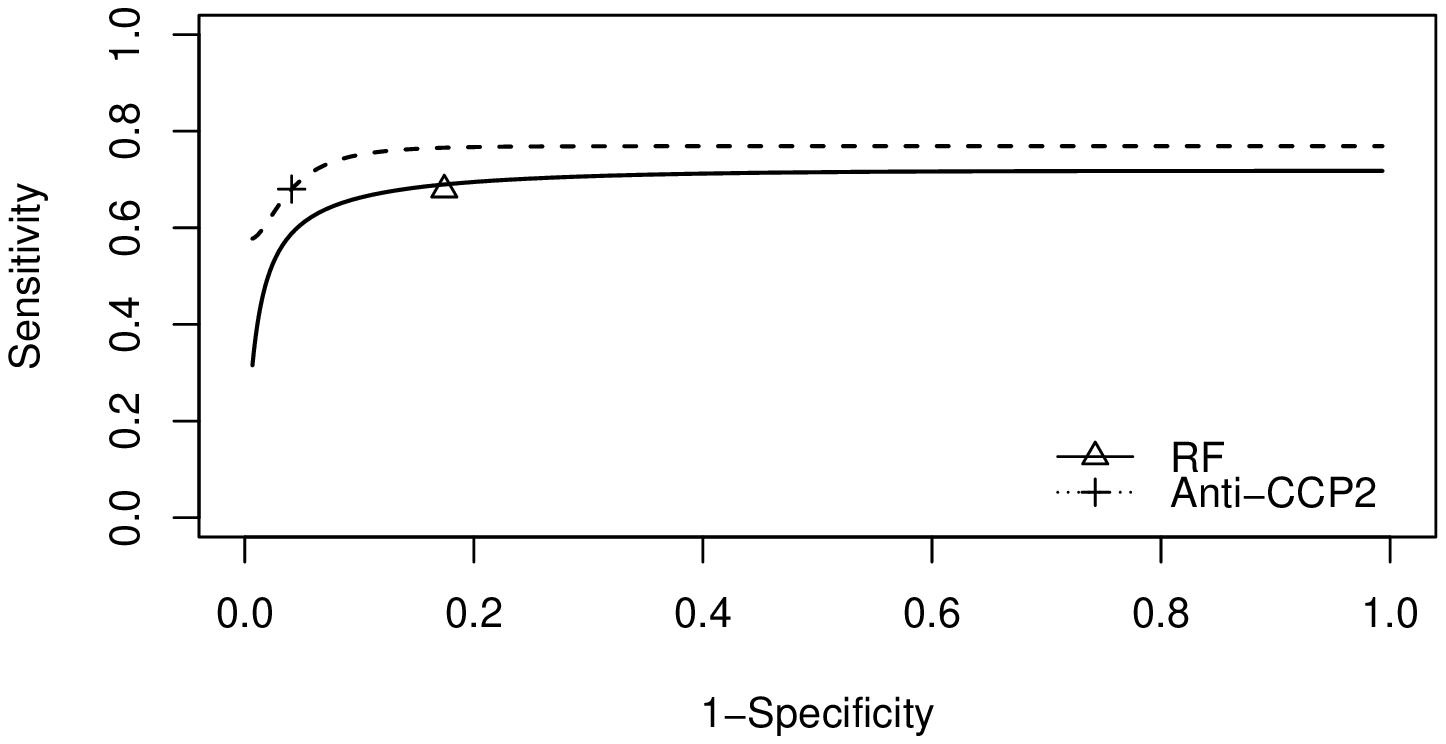}\\\hline
\end{tabular}

\caption{\label{ROC-comp-arthritis}Median regression curves for each test backtransformed to the original scale of sensitivity and specificity for the arthritis data.}
\end{center}
\vspace{-0.5cm}
\end{figure}

Furthermore, a  quadrivariate copula mixed model leads to better inferences than two bivariate copula mixed models  with independence between the two diagnostic tests, since the likelihood has been improved by $10.1=-319.4-(-329.5)$. This indicates that there is strong evidence of  dependence between the two diagnostic tests.  The fact that the best-fitting bivariate copula for the first diagnostic test  is Clayton rotated by 270  degrees (instead of say, BVN) indicates that there is also negative tail dependence (see Figure \ref{ROC-arthritis}). In Figure \ref{ROC-comp-arthritis} is revealed the anti-CCP2 antibody is better compared with RF. 

It is also revealed that the HK method leads to biased estimates for the meta-analytic parameters, their variabilities and Kendall's tau associations as the individual discrete probabilities are large and the margins of the random effects are misspecified (the HK method restricts itself to beta margins).

\subsection{\label{diabetes-section}Diabetes}
The diagnostic tests under investigation are glycated HbA$_{1c}$ and FPG \cite{Bennett-etal2007,Kodama-etal2013}.
Both diagnostic tests are potential alternatives to the oral glucose tolerance test (OGTT), which is commonly used as the gold standard. While the OGTT and FPG require that probands refrain from eating and drinking any liquids, this is not mandatory for the measurement of HbA1c. \cite{hoyer&kuss-2017-sim} Furthermore, HbA1c and FPG are less expensive than the OGTT. \cite{hoyer&kuss-2017-sim} The two systematic reviews include in total 38 different studies.

The log-likelihood showed that a D-vine copula mixed model with Clayton rotated by 90 degrees copula with  normal margins  for test 1 and beta  margins for test 2,  to join the (latent) sensitivities and specificities,  provides the best fit (Table \ref{diabetes-res}).

\begin{table}[!h]
  \centering

  \caption{\label{diabetes-res}Maximized ML and HK log-likelihoods, estimates and standard errors (SE) of the D-vine copula mixed models for the diabetes data.}
    \begin{tabular}{c|cc|cc|c|cc|c|cc}
    \toprule
test&$^\P$ copula&margin&$^\S$ copula&margin&test&$^\dag$ copula&margin&test&$^\ddag$ copula&margin    
    \\\hline
 HbA$_{1c}$& Cln90$^\circ$ &normal&Cln90$^\circ$ & normal& HbA$_{1c}$& BVN&normal&HbA$_{1c}$& Cln90$^\circ$& beta\\
 
FPG& Cln90$^\circ$ &beta&BVN  & beta& FPG& BVN&normal&FPG& Cln90$^\circ$& beta\\

    \midrule
    Param. & Estimate  & SE    & Estimate  & SE    & Param. & Estimate   & SE    & Param. & Estimate   & SE \\\hline
    $\pi_1$ & 0.726 & 0.024 & 0.731 & 0.024 & $\pi_1$ & 0.724 & 0.024 & $\pi_1$ & 0.706 & 0.023 \\
    $\pi_2$ & 0.810 & 0.019 & 0.808 & 0.013 & $\pi_2$ & 0.807 & 0.014 & $\pi_2$ & 0.779 & 0.021 \\
    $\pi_3$ & 0.703 & 0.029 & 0.689 & 0.025 & $\pi_3$ & 0.737 & 0.029 & $\pi_3$ & 0.691 & 0.029 \\
    $\pi_4$ & 0.804 & 0.022 & 0.803 & 0.013 & $\pi_4$ & 0.839 & 0.012 & $\pi_4$ & 0.804 & 0.022 \\
    $\s_1$ & 0.748 & 0.087 & 0.748 & 0.092 & $\s_1$ & 0.762 & 0.099 & $\g_1$ & 0.081 & 0.018 \\
    $\s_2$ & 0.825 & 0.088 & 0.836 & 0.068 & $\s_2$ & 0.846 & 0.076 & $\g_2$ & 0.095 & 0.017 \\
    $\g_3$ & 0.137 & 0.027 & 0.133 & 0.022 & $\s_3$ & 1.049 & 0.112 & $\g_3$ & 0.143 & 0.033 \\
    $\g_4$ & 0.124 & 0.023 & 0.125 & 0.021 & $\s_4$ & 1.055 & 0.118 & $\g_4$ & 0.112 & 0.023 \\
    $\tau_{12}$ & -0.307 & 0.033 & -0.287 & 0.021 & $\tau_{12}$ & -0.378 & 0.060 & $\tau_{12}$ & -0.210 & 0.078 \\
    $\tau_{23}$ & 0.090 & 0.094 & -     & -     & $\tau_{23}$ & 0.016 & 0.094 & $\tau_{23}$ & 0.053 & 0.100 \\
    $\tau_{34}$ & -0.282 & 0.094 & -0.401 & 0.053 & $\tau_{34}$ & -0.401 & 0.072 & $\tau_{34}$ & -0.245 & 0.115 \\
    $\tau_{13|2}$ & 0.375 & 0.080 & -     & -     & $\tau_{13|2}$ & 0.319 & 0.083 & $\tau_{13|2}$ & 0.247 & 0.094 \\
    $\tau_{24|3}$ & 0.400 & 0.085 & -     & -     & $\tau_{24|3}$ & 0.289 & 0.091 & $\tau_{24|3}$ & 0.236 & 0.071 \\
    $\tau_{14|23}$ & 0.223 & 0.112 & -     & -     & $\tau_{14|23}$ & 0.318 & 0.072 & $\tau_{14|23}$ & 0.156 & 0.080 \\\hline
    $-\log L$ & \multicolumn{2}{c}{772.6} & \multicolumn{2}{c}{783.4} & $-\ell$ & \multicolumn{2}{c}{779.9} & $-\ell$ & \multicolumn{2}{c}{778.5} \\
    \bottomrule
    \end{tabular}%
 \vspace{-1ex} 
\begin{flushleft}
\begin{footnotesize}
Cln$90^\circ$: Clayton rotated by 90 degrees;
$\P$: Best fit; $^\S$: Separate bivariate meta-analysis for each test;
$^\dag$: The resulting model is the same as the quadrivariate  GLMM; 
$^\ddag$: HK approximated likelihood method. 
\end{footnotesize} 
\end{flushleft} 
\end{table}

\begin{figure}[!h]
\begin{center}
\begin{footnotesize}
\begin{tabular}{|cc|}
\hline
HbA$_{1c}$ & FPG\\\hline

\includegraphics[width=0.5\textwidth]{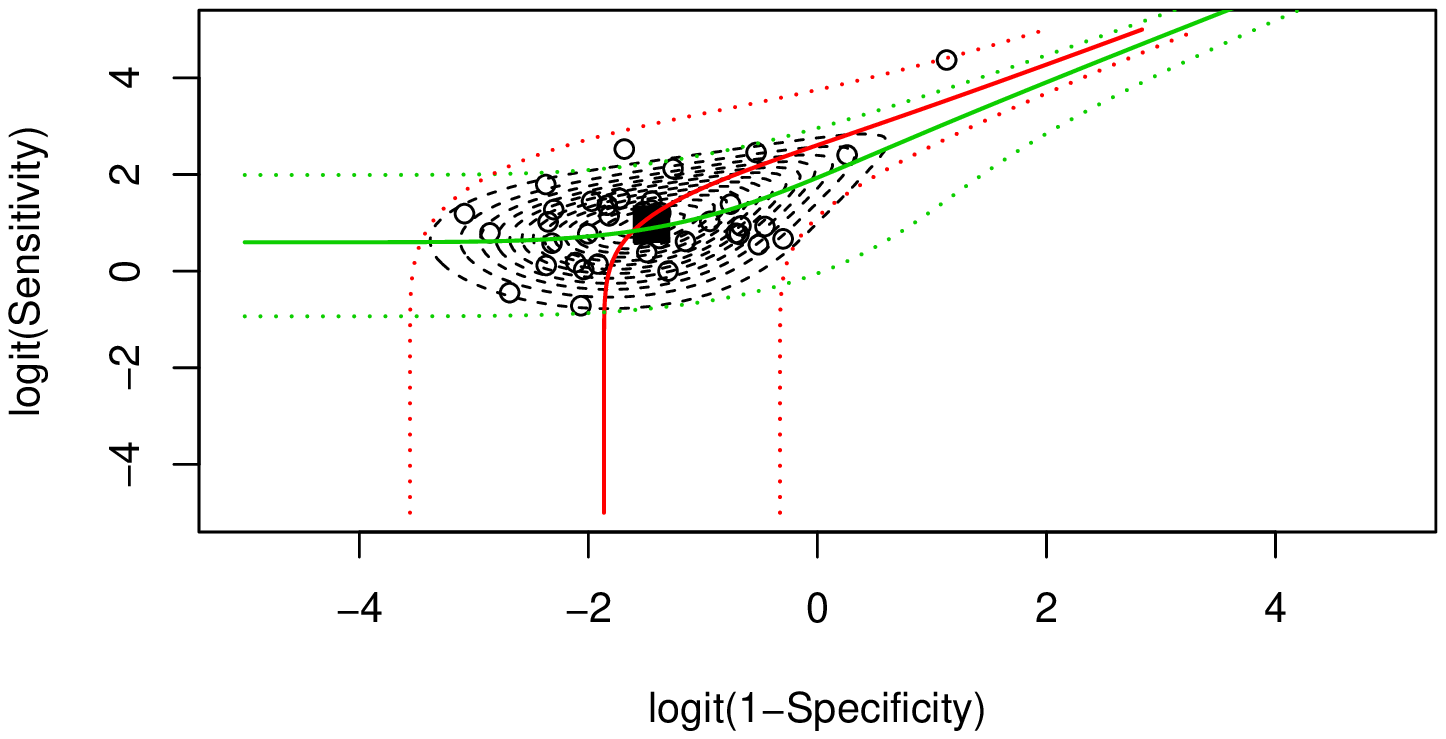}
&

\includegraphics[width=0.5\textwidth]{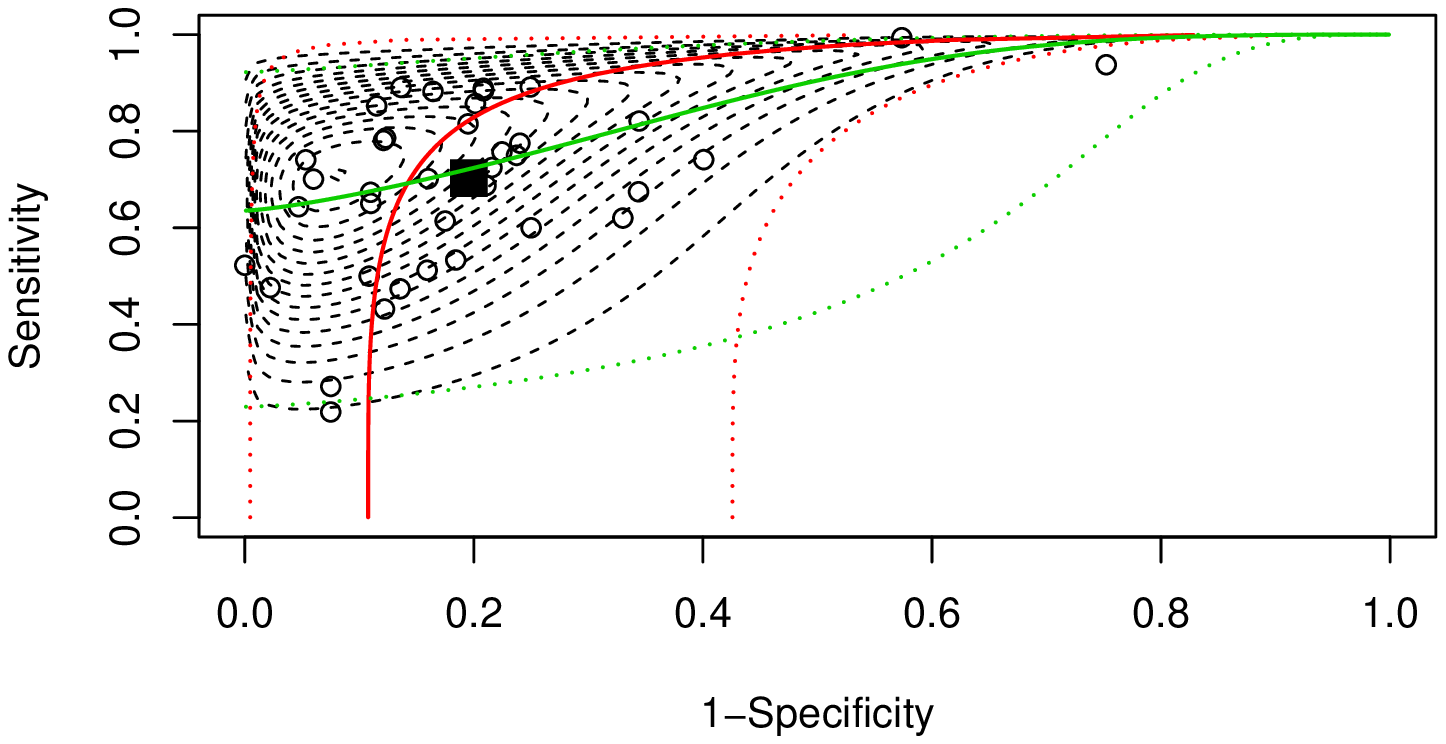}\\\hline
\end{tabular}
\end{footnotesize}
\caption{\label{ROC-diabetes}Contour plots (predictive region)  and quantile  regression curves  from the best fitted D-vine copula mixed model for the diabetes data. Red and green lines represent the quantile  regression curves $x_1:=\widetilde{x}_1(x_2,q)$ and $x_2:=\widetilde{x}_2(x_1,q)$, respectively; for $q=0.5$ solid lines and for $q\in\{0.01,0.99\}$ dotted lines (confidence region).  median regression curve for each model. In case of HbA$_{1c}$ the axes are in  logit scale since  we also plot  the estimated contour plot of the random effects distribution as predictive region; this has been estimated for the logit pair of (Sensitivity, Specificity).}
\end{center}
\vspace{-1cm}
\end{figure}

\begin{figure}[!h]
\begin{center}
\begin{tiny}
\psfrag{a}{HbA$_{1c}$}\psfrag{b}{FPG}
\begin{tabular}{|cc|}
\hline
$x_1:=\widetilde{x}_1(x_2,q=0.5)$ & $x_2:=\widetilde{x}_2(x_1,q=0.5)$\\\hline

\includegraphics[width=0.5\textwidth]{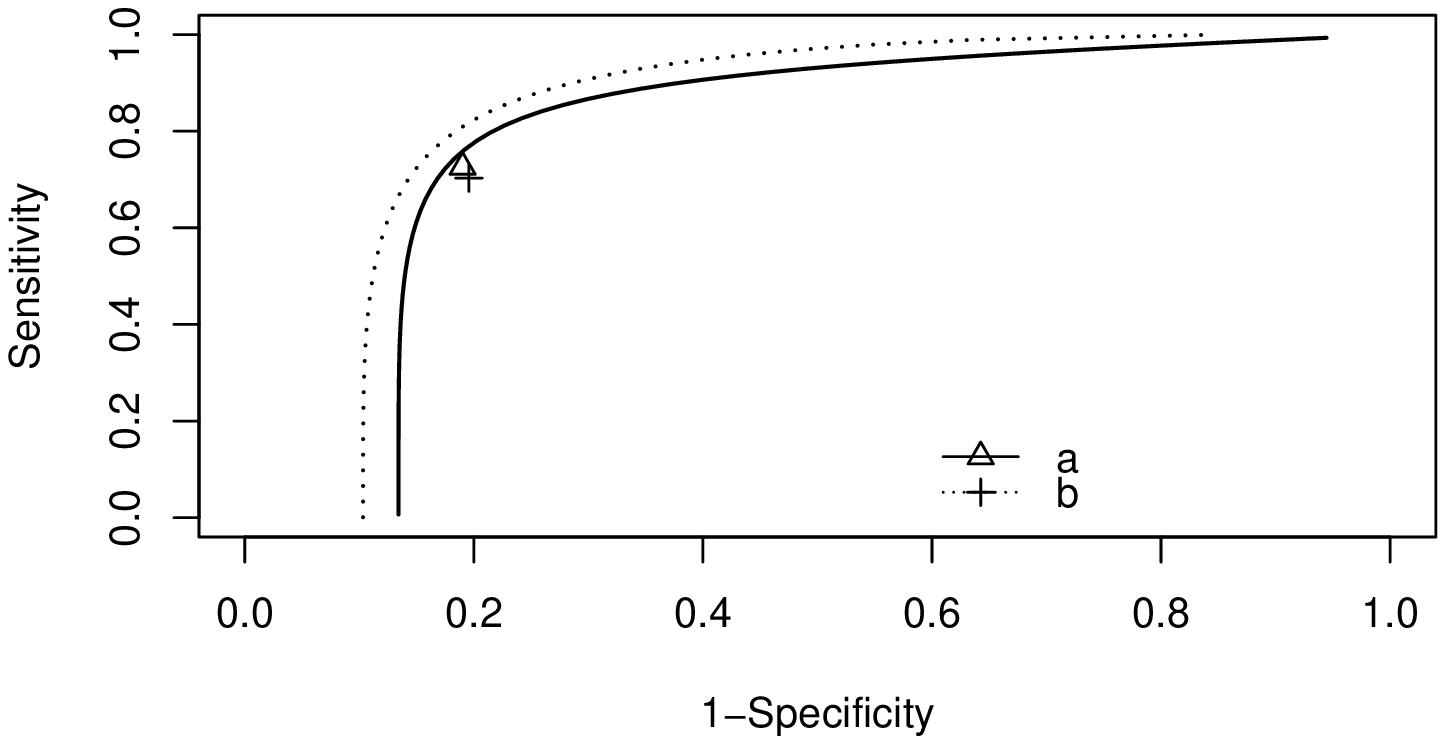}
&

\includegraphics[width=0.5\textwidth]{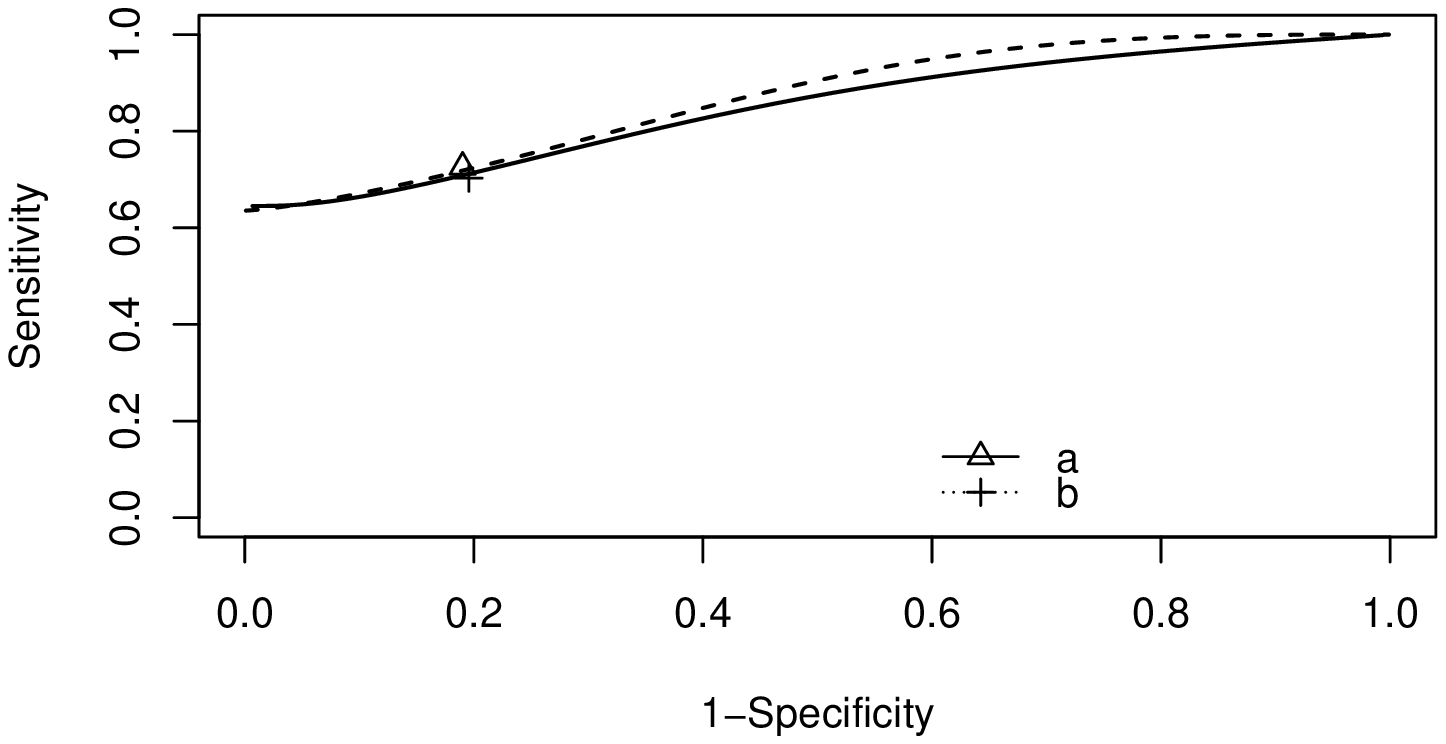}\\\hline
\end{tabular}
\end{tiny}
\caption{\label{ROC-comp-diabetes}Median regression curves for each test back-transformed to the original scale of sensitivity and specificity for the diabetes data.}
\end{center}
\vspace{-0.5cm}
\end{figure}

Furthermore, a quadrivariate copula mixed model leads to better inferences than two bivariate copula mixed models  with independence between the two diagnostic tests since the likelihood has been improved by $10.8=-772.6-(-783.4)$. This indicates that there is strong evidence of  dependence between the two diagnostic tests.  The fact that the best-fitting bivariate copula for both diagnostic tests is Clayton rotated by 90 (instead of say, BVN) indicates that there is also negative tail dependence (see Figure \ref{ROC-diabetes}). In Figure \ref{ROC-comp-diabetes} is revealed the FPG antibody is slightly better compared with HbA$_{1c}$.

It is also demonstrated  that the HK method leads to biased estimates for the meta-analytic parameters, as although   the individual discrete probabilities are small,  the margins of the random effects are misspecified (the HK method restricts itself to beta margins).

\section{\label{discussion}Discussion}

We have proposed a D-vine copula mixed model for joint meta-analysis and comparison of  diagnostic tests. This is the most general meta-analytic model, with univariate parameters separated from dependence parameters. Our general model includes the quadrivariate GLMM as a special case and can provide an improvement over the latter based on log-likelihood and thus, can provide a better statistical inference for the SROC and the meta-analytic parameters of interest.

For the quadrivariate D-vine copula mixed model the model parameters (including dependence parameters), the choice of the bivariate copulas to link the sensitivity and specificity, and the choice of the margin  for each test affect the shape of the SROC curves.  
The HK\cite{hoyer&kuss-2017-sim} approximation method   cannot be used to produce the SROC curves, since the  dependence parameters  affect the shape of the SROC curve and these are generally underestimated.   Note in passing that the quadrivariate GLMM \cite{hoyer&kuss-2016-smmr} can produce SROC curves  but these are restricted to the elliptical shape.

We have proposed a numerically stable ML estimation technique based on Gauss-Legendre quadrature; the crucial step is to convert from independent to dependent quadrature points.  Although there is an issue of computational burden as the number of quadrature points increase, this will subside, as computing technology is advancing rapidly. 
Any comparison with the HK method in terms of computing time is a digression. It is obvious that the HK method is much faster then the ML method (even when $n_q=15$), since a numerically more difficult 4-dimensional integral calculation is replaced with a much simpler computationally copula density value. However, theoretically there are still problems for large individual probabilities, since the HK approximation of `large steps' in the cdf is poor. In fact, we novelty propose simple diagnostics (descriptive statistics such as a histogram) to judge if the HK method is reliable and reduce by its use the computational burden when it is possible (i.e., for small individual probabilities). Nevertheless,  even at this case the HK method might lead to biased estimates since it assumes  an Beta$(\pi,\gamma)$ distribution for the marginal modelling of
the latent proportions. This was the case for the diabetes data in Section \ref{diabetes-section}.

\section*{Software}
{\tt R} functions to  implement the D-vine copula mixed model for meta-analysis  and comparison of two diagnostic tests are  part of the  {\tt  R} package {\tt  CopulaREMADA}. \cite{Nikoloulopoulos-2015}

\section*{Acknowledgements}
The simulations presented in this paper were carried out on the High Performance Computing Cluster supported by the Research and Specialist Computing Support service at the University of East Anglia.

\end{document}